\documentclass{aa}

\usepackage{graphicx}
\usepackage{txfonts}
\usepackage{color}

\begin{document}

\title{Tidal evolution of galaxies in the most massive cluster\\ of IllustrisTNG-100}

\author{Ewa L. {\L}okas
}

\institute{Nicolaus Copernicus Astronomical Center, Polish Academy of Sciences,
Bartycka 18, 00-716 Warsaw, Poland\\
\email{lokas@camk.edu.pl}}


\abstract{
We study the tidal evolution of galaxies in the most massive cluster of the IllustrisTNG-100 simulation. For the
purpose of this work, we selected 112 galaxies with the largest stellar masses at present and followed their properties over
time. Using their orbital history, we divided the sample into unevolved (infalling), weakly evolved (with one pericenter
passage), and strongly evolved (with multiple pericenters). The samples are clearly separated by the value of the
integrated tidal force from the cluster the galaxies experienced during their entire evolution and their properties
depend strongly on this quantity. As a result of tidal stripping, the galaxies of the weakly evolved sample lost
between 10 and 80\% of their dark mass and less than 10\% of stars, while those in the strongly evolved one lost more than
70\% of dark mass and between 10 and 55\% of stellar mass, and are significantly less, or not at all dark-matter dominated.
While 33\% of the infalling galaxies do not contain any gas, this fraction increases to 67\% for the weakly evolved sample, and
to 100\% for the strongly evolved sample. The strongly evolved galaxies lose their gas earlier and faster (within 2-6
Gyr), but the process can take up to 4 Gyr from the first pericenter passage. These galaxies are redder and more metal
rich, and at redshift $z=0.5,$ the population of galaxies in the cluster becomes predominantly red. As a result of tidal
stirring, the morphology of the galaxies evolves from oblate to prolate and their rotation is diminished, thus the
morphology-density relation is reproduced in the simulated cluster. The strongly evolved sample contains at least six
convincing examples of tidally induced bars and six more galaxies that had their bars enhanced by their interaction
with the cluster.
}


\maketitle

\section{Introduction}

Evolution of galaxies in clusters has been the subject of intensive study for five decades. The early findings
were summarized in a review by \citet{Dressler1984}, who was then already able to identify many of the processes
involved. Clusters of galaxies are presently the largest gravitationally bound structures in the Universe. According to
the commonly accepted hierarchical theory of structure formation, they grow through the accretion of smaller structures such as
galaxies and groups. As new galaxies enter the cluster, they start to interact with it via a number of processes that
affect their properties. Since the infalling galaxies increase their velocities and continue to move rather fast as
they orbit in the cluster, their interactions with other galaxies become less probable. Instead, the effect of the
cluster as a whole becomes dominant.

The main processes that can affect galaxies in clusters fall into two categories: gravitational and hydrodynamical.
The gravitational effects manifest themselves mainly as tidal forces exerted by the cluster gravitational potential on
the galaxies. The tidal effects result in tidal stripping of the galaxies in the form of mass loss and tidal stirring
that disturbs the morphology and dynamics of the galaxies, and that may lead to the transformation of their shapes from
disks to spheroids and the rotation of the stars to random motions \citep{Valluri1993, Moore1999, Gnedin2003,
Mastropietro2005}. The effects of tidal stirring manifest themselves in the morphology-density relation: the early-type objects tend to dominate the cluster center, while late-type ones are predominantly found in the outskirts of
clusters \citep{Dressler1980}. Such a relation has been observed in many clusters including the well-studied Coma
cluster where ellipticals are found to have a significantly steeper density profile than spirals \citep{Lokas2003}.

Among the hydrodynamical processes, the one of ram pressure stripping by the hot intracluster gas seems to play a
crucial role \citep{Gunn1972}. As a normal spiral galaxy with a significant gas content enters the cluster, it is swept
clean of the interstellar medium, and the lost gas feeds the central giant galaxy or is heated to the temperature of the
intracluster gas and distributed smoothly within the cluster. The loss of gas leads to reduced star formation in
cluster galaxies compared to the field, a phenomenon often referred to as environmental quenching \citep{Lotz2019}. As
a result, galaxy clusters at higher redshift contain a larger fraction of blue galaxies than the low-redshift ones,
which has long been known as the Butcher-Oemler effect \citep{Butcher1978, Butcher1984}. The reality of the
ram pressure stripping has now been confirmed by numerous observations of "jellyfish" galaxies in clusters
\citep{Ebeling2014, McPartland2016, Poggianti2016}.

Until recently, theoretical studies of galaxy evolution in clusters using $N$-body and hydro simulations were plagued
by the difficulty to include all physical processes operating on different scales and were thus necessarily restricted
to either pure dark matter simulations, semi-analytic models, low-resolution, or controlled simulations of single
galaxies evolving in the cluster potential. The suite of Illustris and the follow-up IllustrisTNG simulations
\citep{Springel2018, Marinacci2018, Naiman2018, Nelson2018, Pillepich2018a} are the first successful attempts to model
the formation of galaxies in the cosmological context, including their evolution in clusters. On one hand, the use of
large enough simulation boxes makes it possible to follow structure formation from cosmologically motivated initial conditions and
provides a realistic environment of galaxy formation, while on the other hand, high enough resolution makes it possible to study the
evolution of properties of individual galaxies. Most importantly, the simulations include state-of-the-art treatment of
baryonic processes, such as gas radiative mechanisms, star formation, stellar population evolution and chemical
enrichment, stellar and black hole feedback, and the evolution of cosmic magnetic fields \citep{Pillepich2018b,
Weinberger2017}.

The IllustrisTNG simulations have already been used in a number of studies pertaining to galaxy clusters.
\citet{Pillepich2018a} looked at the stellar content of clusters, \citet{Gupta2018} studied chemical pre-processing of
cluster galaxies, \citet{Barnes2018} investigated the census of cool-core galaxy clusters, \citet{Yun2019} focused on
gas-stripping phenomena and jellyfish galaxies, while \citet{Sales2020} studied the formation of ultra-diffuse galaxies
in clusters.

This work is to some extent motivated by the study of \citet{Lokas2016}, who used controlled $N$-body simulations to
follow the evolution of a single, initially disky galaxy similar to the Milky Way on different orbits around a cluster
similar to Virgo. They found that with subsequent pericenters, the galaxy tends to lose rotation and increase the amount
of random motions of the stars while its disk thickens. The tidal forces from the cluster were also found to be very
effective in the formation of tidally induced bars and spiral arms \citep{Semczuk2017}. The mechanisms operating in the
cluster are essentially the same as those in the Local Group, where tidal stirring has been invoked as a way to
transform disky dwarf galaxies orbiting the Milky Way or M31 into dwarf spheroidals \citep{Mayer2001, Kazantzidis2011}.
In this case, the formation of tidally induced bars was also found to be an intermediate stage in the transformation
\citep{Lokas2014, Gajda2017}. These configurations are just two among many that are known to lead to the
formation of tidally induced bars, including fly-by interactions of galaxies of similar size or perturbations by
satellites \citep{Gerin1990, Noguchi1996, Miwa1998, Mayer2004, Berentzen2004, Lang2014}.

In this paper, we attempt to place this tidal evolution scenario in the cosmological context by studying the evolution of
galaxies in the most massive cluster of the Illustris TNG-100 simulation. We take advantage not only of the realistic
treatment of the evolving environment but also the inclusion of baryonic processes. We restrict the analysis to about a
hundred galaxies with the highest stellar mass, which are bound to the cluster at the present epoch. Such a choice
allows us on one hand to study the evolution of the structural properties of individual galaxies with sufficient
accuracy, and on the other to have adequate samples of galaxies in different evolutionary stages for comparison.

The paper is organized as follows. The main results of this work are presented in Section~2. In the subsequent
subsections of this part, we present our sample selection, describe the orbital evolution of the galaxies, the mass loss
they experience, and we estimate different timescales of their evolution. Next, we focus on the evolution of color and
metallicity, as well as the morphological and kinematical changes induced in the galaxies by the cluster, and we discuss the
properties of tidally induced bars. The summary and discussion follow in Section~3.

\section{Evolution of the galaxies}

\begin{figure*}[ht!]
\centering
\includegraphics[width=18cm]{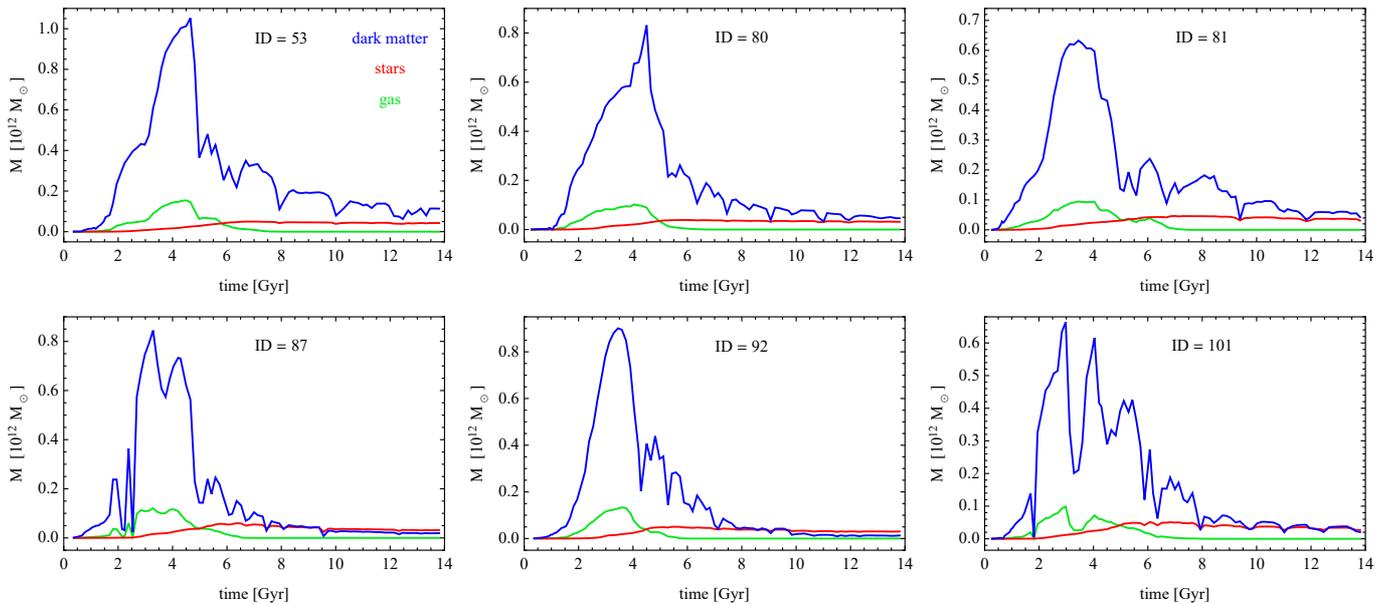}
\caption{Evolution of masses of galaxies with ID=53, 80, 81, 87, 92, and 101 in the simulated cluster.
The mass of the dark, stellar, and gas component is shown with the blue, red, and green lines,
respectively, with the mass scale adjusted to the range of dark matter masses.}
\label{evolution_mass}
\end{figure*}

\begin{figure*}[ht!]
\centering
\includegraphics[width=6cm]{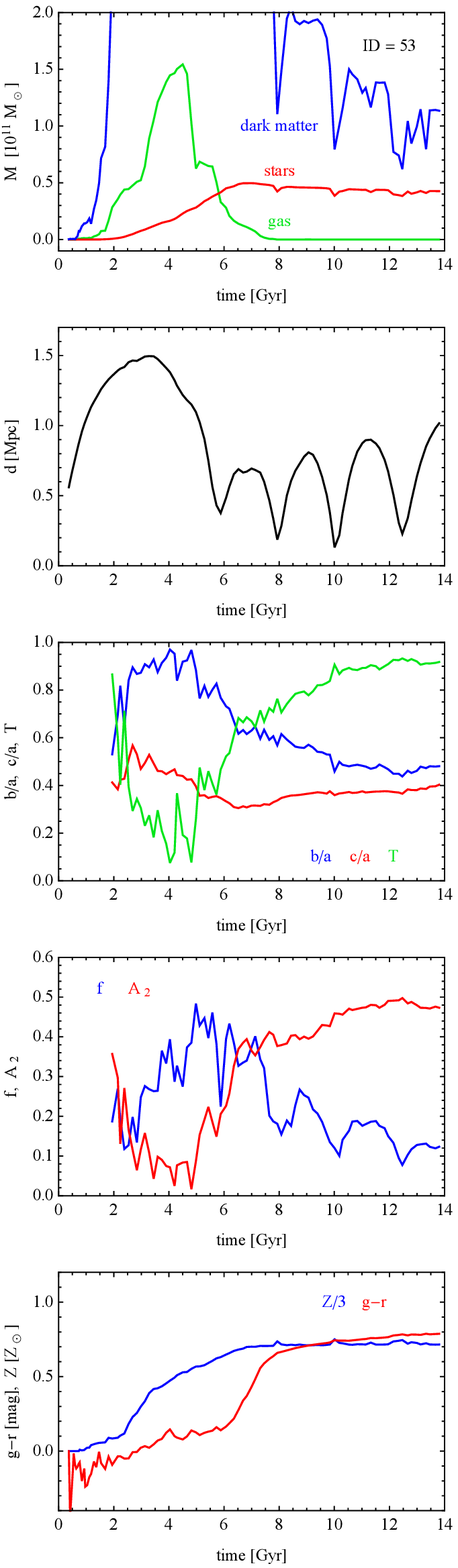}
\includegraphics[width=6cm]{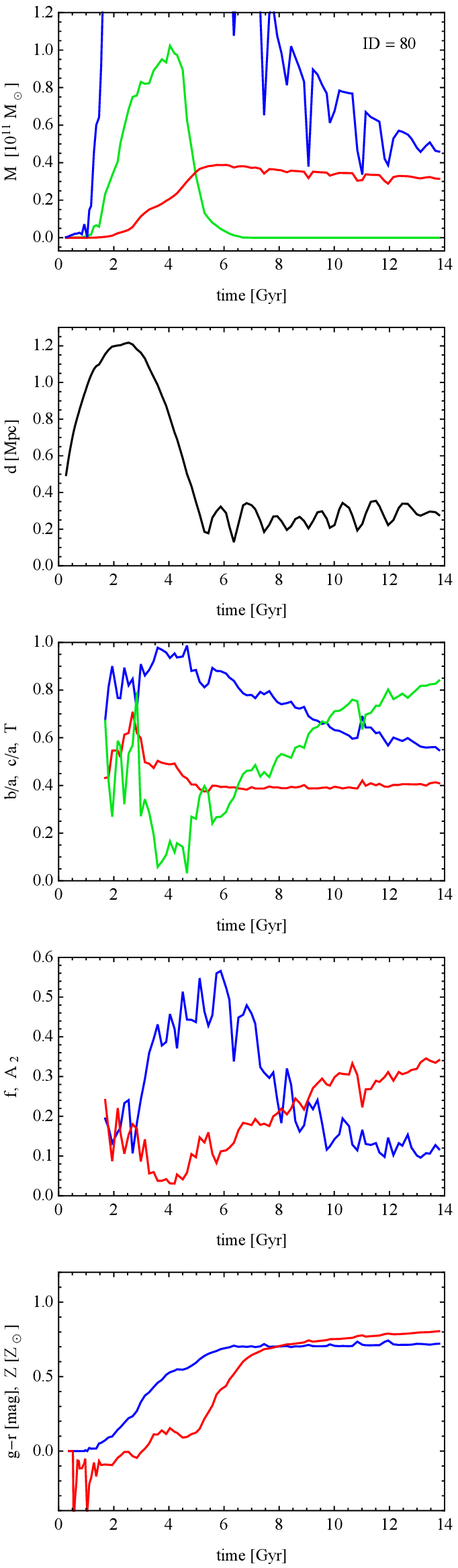}
\includegraphics[width=6cm]{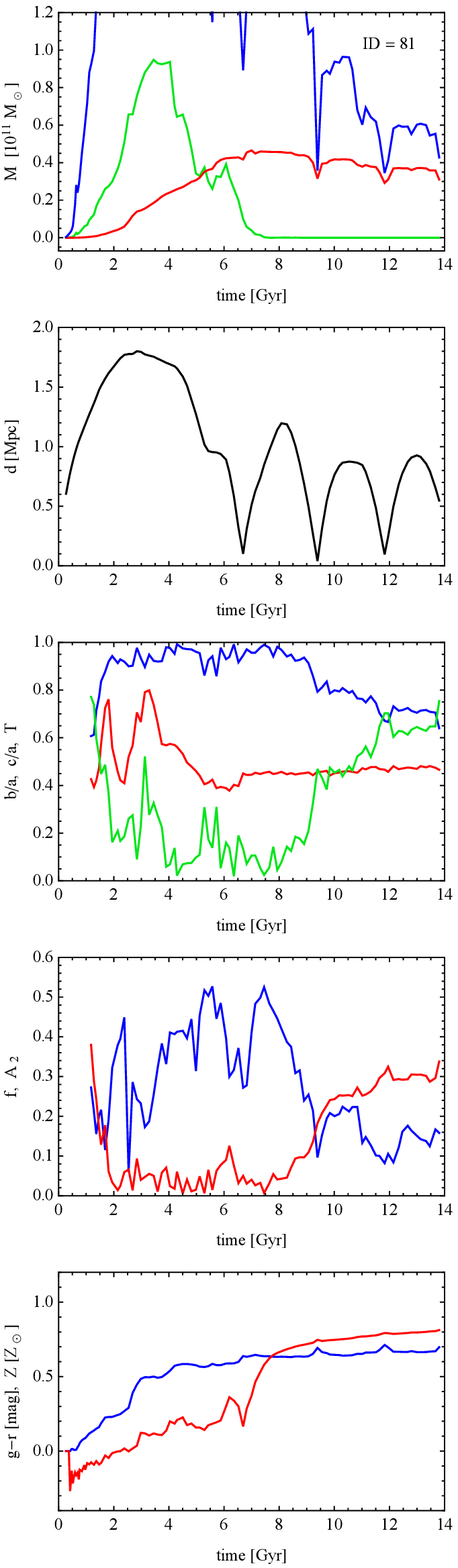}
\caption{Evolution of galaxies in simulated cluster. Columns: results for different galaxies, with ID=53, 80
and 81. Upper row: the evolution of the dark, stellar and gas mass shown with the blue, red, and green lines,
respectively. The data are the same as in Fig.~\ref{evolution_mass}, but the mass scale was adjusted to the range of
stellar and gas masses. Second row: the orbit of the galaxy within the cluster in terms of its distance from the
cluster center in physical units. Third row: the evolution of three structural properties of the galaxies, the axis
ratios $b/a$ (blue line) and $c/a$ (red), and the triaxiality parameter $T$ (green). Fourth row: the rotation measure
in terms of the fractional mass of stars on circular orbits $f$ (blue) and the bar strength $A_2$ (red). Fifth row: the
$g-r$ color and metallicity $Z$ (in solar units, divided by 3 to fit the same scale as the color).} \label{evolution1}
\end{figure*}

\begin{figure*}
\centering
\includegraphics[width=6cm]{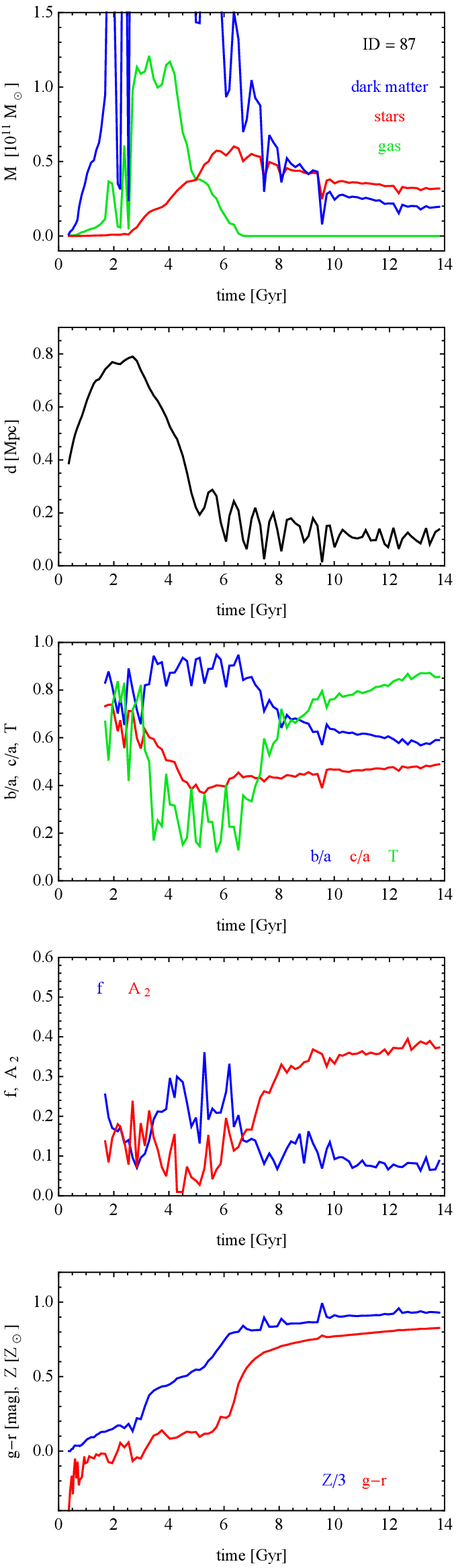}
\includegraphics[width=6cm]{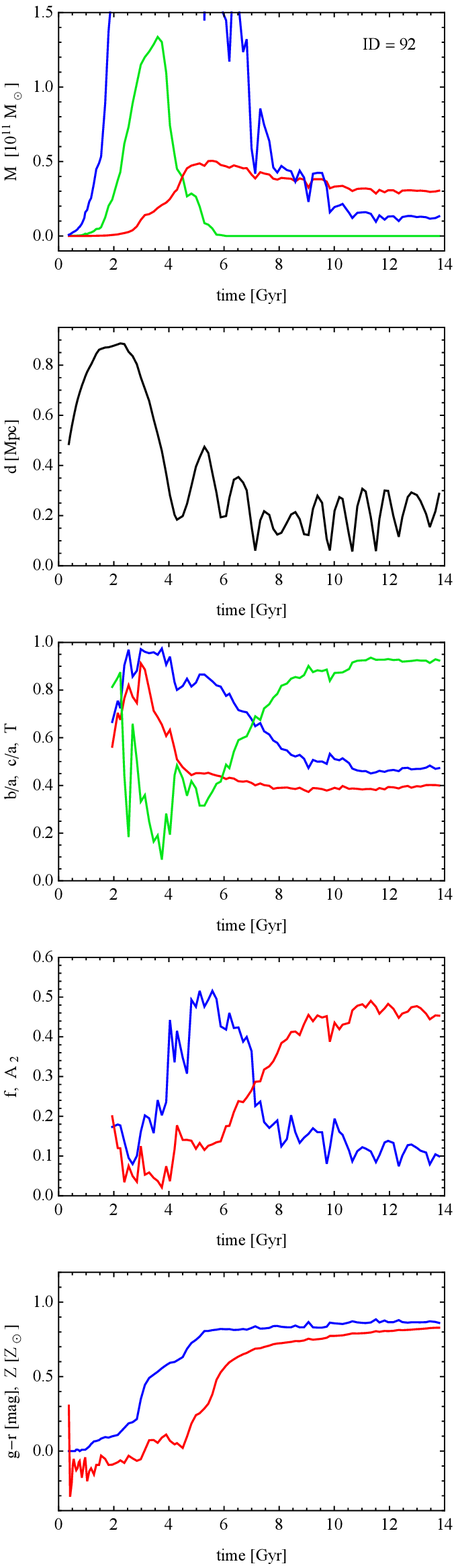}
\includegraphics[width=6cm]{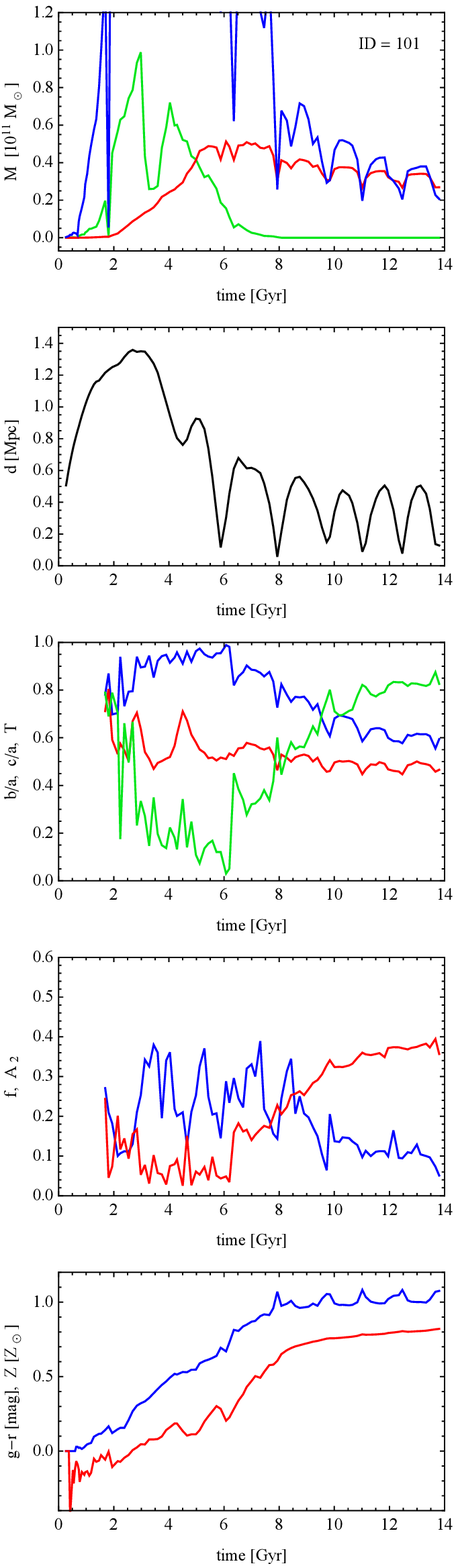}
\caption{Same as Fig.~\ref{evolution1}, but for galaxies with ID=87, 92 and 101.}
\label{evolution2}
\end{figure*}

\subsection{Sample selection}

In this work, we used the publicly available simulation data from IllustrisTNG described by \citet{Nelson2019}. For
the purpose of this study, we selected the simulation TNG-100 from the IllustrisTNG set, where the number 100 refers to
the size of the simulation box in Mpc. The simulation data are available in 100 snapshots from redshift $z=20$ to the
present epoch, $z=0$, which is sufficient to trace the evolution of galaxies. The simulation box contains two
massive galaxy clusters at $z=0$ from which we select the most massive one (with total mass $M = 4 \times 10^{14}$
M$_{\odot}$), which is more regular and relaxed as indicated by its temperature distribution. This object bears the
identification number ID=0 in the catalog of halos and subhalos identified in this simulation at $z=0$.

Using the simulation output 99 corresponding to $z=0,$ we selected subhalos identified with the Subfind
algorithm \citep{Springel2001} parented by this halo with stellar masses $M_* > 10^{10}$ M$_\odot$. This threshold
translates to the number of stars per galaxy above $10^4,$ which allows us to reliably measure their global structural
properties. After rejecting spuriously identified subhalos with almost no dark matter, we are left with 112 galaxies
whose properties we study. We trace the history of these galaxies using main progenitor branches of Sublink merger
trees for the subhalos \citep{Rodriguez2015} to the earliest time possible, that is, to the snapshot when the subhalo
was first identified. For the selected galaxies, these earliest snapshots of subhalo existence are among the first few
snapshots available, that is, the galaxies can be traced down to redshifts $z=10-20$.

One of the most interesting properties of galaxies from the point of view of tidal evolution is their mass. Using the
galaxy properties provided in the Sublink catalogs, we constructed the evolutionary histories of the galaxies in
terms of their mass evolution in three components: dark matter, stars and gas. Examples of such evolution for six
galaxies in the cluster are shown in Fig.~\ref{evolution_mass}. The most characteristic feature of these curves is that
all the galaxies increase their dark masses up to the point when they start to lose it after entering the cluster, as
we discuss in detail below.

\subsection{Orbital history}

Using the subhalo positions provided in the Sublink catalogs, we calculated the orbits of the galaxies in the cluster in
terms of their physical distance from the cluster center as a function of time. Examples of the evolution of this
distance for six galaxies on tight orbits around the cluster are shown in the second row of panels in
Figs.~\ref{evolution1} and \ref{evolution2}.

Inspection of the orbital history of the whole sample of 112 galaxies allows us to divide them into three distinct
subsamples. It turns out that 48 out of 112 galaxies (43\%) are on their first infall into the cluster and did not have
any pericenter passage on their orbit. These objects are bound to the cluster and approaching it, but in some cases are
still well beyond the virial radius of the cluster, which at $z=0$ is equal to 2 Mpc. This sample comprises a variety
of galaxies of different types, properties, and histories that may have interacted with other objects or evolved in
groups prior to their accretion by the cluster but were not yet significantly affected by the cluster environment. We do
not study these galaxies further in detail here, but we use them as a reference for comparison with the properties
of galaxies that were affected by the cluster. In the following, we refer to this subsample as our reference or
unevolved sample.

The next, more interesting subsample contains 37 out of 112 galaxies (33\%) that experienced one pericenter passage on
their orbit around the cluster with the pericenter distance within the virial radius of the cluster at that time. We refer to these galaxies as weakly evolved. The most interesting sample from the point of view of the effects of
environment is the one containing 27 out of 112 galaxies (24\%) that experienced more than one pericenter passage. We
consider this subsample as strongly evolved.

In order to quantify the strength of the interaction of each galaxy with the cluster we used the orbital data to
calculate the integrated tidal force (ITF) from the cluster experienced by the galaxy over its entire history. Looking
at the dependence of different galaxy properties on this quantity rather than just redshift or time allows us to
isolate the effect of the cluster from the cosmological evolution of galaxies in general. ITF was obtained by
multiplying the tidal force experienced by the galaxy at a given simulation snapshot by the timestep between this and
the following snapshot, and summing up such terms for all snapshots. The tidal force was approximated as in
\citet{Lokas2011a} by the formula $F_{\rm tidal} \propto r M(<d)/d^3$, where $r$ is the characteristic scale length of
the galaxy at which the tidal force operates, $d$ is the distance between the center of the galaxy and the center of
the cluster, and $M(<d)$ is the mass of the cluster within this distance.

\begin{table*}
\caption{Properties of galaxies identified as tidally induced bars at $z=0$.}
\label{table}
\centering
\begin{tabular}{r c c c r c c c c c c c}
\hline\hline
ID \     & log ITF& $d$   & $M_*$                & $M_{\rm dm}$ \ \ \ \ & $g-r$& $Z$ & $b/a$  & $c/a$ & $T$  & $f$  &
$A_{2,{\rm max}}$ \\
         &        & [Mpc] &[$10^{10}$ M$_\odot$] & [$10^{10}$ M$_\odot$]& [mag]& [Z$_{\odot}]$ &        &       &      &
& \\ \hline
 53 \    & 6.93   & 1.02 &  4.27                 & 11.35 \ \ \ \        & 0.79 & 2.15 & 0.48   & 0.40  & 0.92 & 0.12 &
0.67 \\
 80 \    & 7.41   & 0.28 &  3.14                 &  4.59 \ \ \ \        & 0.80 & 2.16 & 0.55   & 0.41  & 0.84 & 0.12 &
0.58 \\
 81 \    & 7.17   & 0.55 &  3.10                 &  4.28 \ \ \ \        & 0.81 & 2.09 & 0.64   & 0.47  & 0.75 & 0.16 &
0.45 \\
 87 \    & 8.07   & 0.14 &  3.19                 &  1.96 \ \ \ \        & 0.83 & 2.79 & 0.59   & 0.49  & 0.86 & 0.09 &
0.47 \\
 92 \    & 7.70   & 0.29 &  3.04                 &  1.32 \ \ \ \        & 0.83 & 2.58 & 0.47   & 0.40  & 0.92 & 0.10 &
0.65 \\
$101 \:$ & 7.39   & 0.13 &  2.69                 &  2.07 \ \ \ \        & 0.82 & 3.22 & 0.59   & 0.47  & 0.83 & 0.05 &
0.49 \\
\hline
\end{tabular}
\end{table*}

As the characteristic scale of the galaxy, we adopted $r=1$ kpc for all of them, which is always of the order of the
radius containing half stellar mass. The distance $d$ was taken from the orbital histories of the galaxies, such as
those shown in the second row of panels in Figs.~\ref{evolution1} and \ref{evolution2}. The mass of the cluster $M(<d)$
was calculated for each snapshot using the virial masses provided by the Sublink tree catalog for the cluster and
approximating the mass dependence on radius by the Navarro-Frenk-White profile \citep{Navarro1997,
Lokas2001}. The integrated tidal force thus calculated is in units of M$_\odot$ Gyr kpc$^{-2}$, but since its values are
large, in the following we use $\log$ ITF without quoting the units, since only the relative differences between its
values for different galaxies are important. The values of $\log$ ITF for our selection of six galaxies in
Figs.~\ref{evolution1} and \ref{evolution2} are listed in the second column of Table~\ref{table}, and their present
distances from the cluster center in the third column.

\subsection{Mass loss}

The evolution of different mass components in time for the six selected galaxies shown in
Fig.~\ref{evolution_mass} is replotted in the upper rows of panels in Figs.~\ref{evolution1}
and \ref{evolution2} with the mass scale adjusted to the mass of gas and stars, to enable easy comparison with the
orbital history and other properties of the galaxies. The stellar and dark masses of these galaxies in the final
simulation snapshot, at $z=0$, are listed in the fourth and fifth columns of Table~\ref{table}.

The dark versus stellar masses of the 112 galaxies from our sample at $z=0$ are shown in the upper panel of
Fig.~\ref{masses} with points of different color indicating their orbital history. Here, and in the following Figures,
the galaxies that are infalling and did not yet have any pericenter passage are denoted with blue circles, those with a single pericenter with green dots, and those with multiple pericenters with red symbols. As expected, for the majority
of objects dark matter masses are higher than the stellar ones. However, the dark masses are systematically much higher
than the stellar ones for the unevolved and weakly evolved sample (blue and green points) than for the strongly evolved
one (red symbols). This already suggests that the strongly evolved sample lost much more dark matter than the other
samples.

In the lower panel of Fig.~\ref{masses}, we plot the ratio of the dark to stellar mass for each galaxy as a function of
the ITF. We see that the samples are now clearly separated by the total tidal force experienced by the galaxies. The
division between the unevolved and weakly evolved samples (blue and green points) occurs around log ITF = 5.5 and
between the weakly and strongly evolved samples (green and red points) at $\log$ ITF = 6.5. In addition, the weakly and
strongly evolved samples (green and red) form a distinct branch with a strong dependence of the mass ratio on ITF.
The correlation between the two quantities is almost linear between the log of mass ratio and log ITF. Along this
branch, the galaxies from the weakly evolved sample (green) have dark to stellar mass ratios roughly between 5 and 20,
while the strongly evolved objects only between 0.4 and 6.

\begin{figure}
\centering
\includegraphics[width=7.1cm]{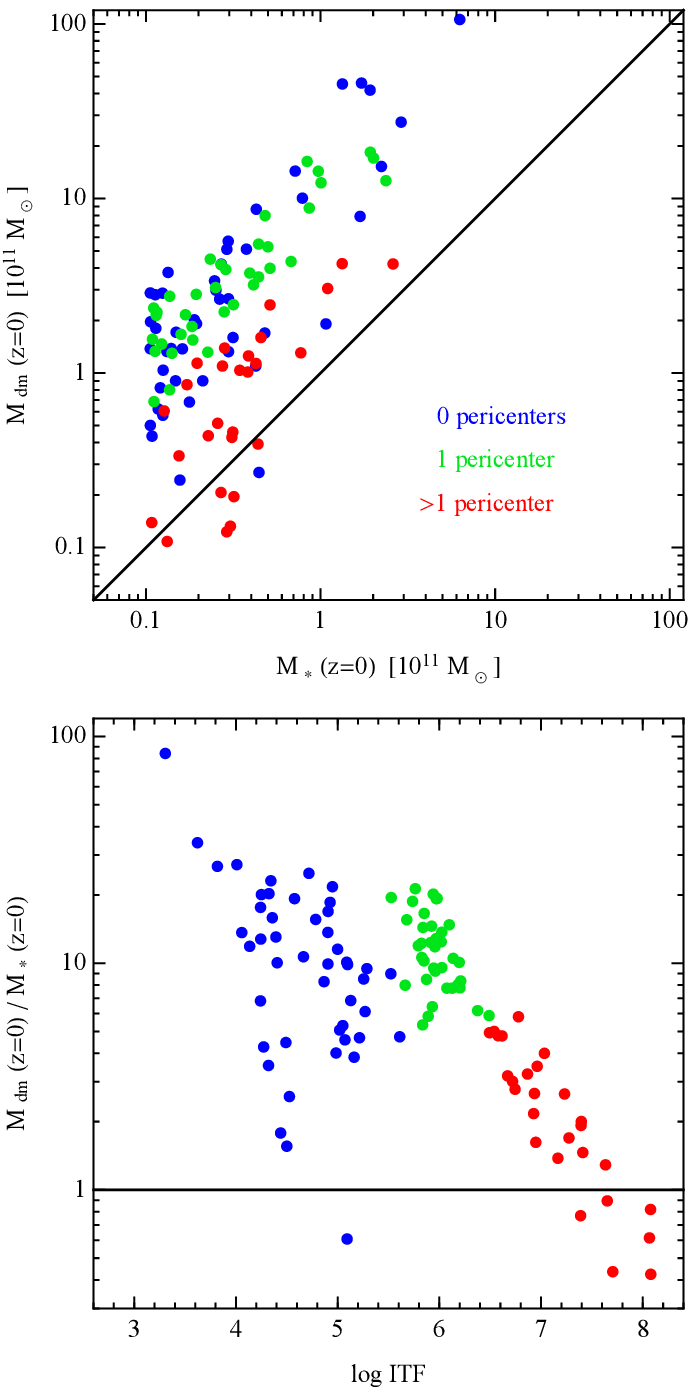}
\caption{Stellar and dark matter masses of selected galaxies at the last simulation snapshot corresponding to the
present time. Upper panel: the dark versus stellar masses for our selection of 112 galaxies with colors coding
their orbital history. Galaxies infalling for the first time are marked with blue symbols, those that experienced one
pericenter passage with green dots and those with more than one pericenter passage with red points. Lower panel:
the ratio of dark to stellar mass as a function of the integrated tidal force. In both panels, the black line
indicates the equality of dark and stellar masses.}
\label{masses}
\end{figure}

\begin{figure}
\centering
\includegraphics[width=7.1cm]{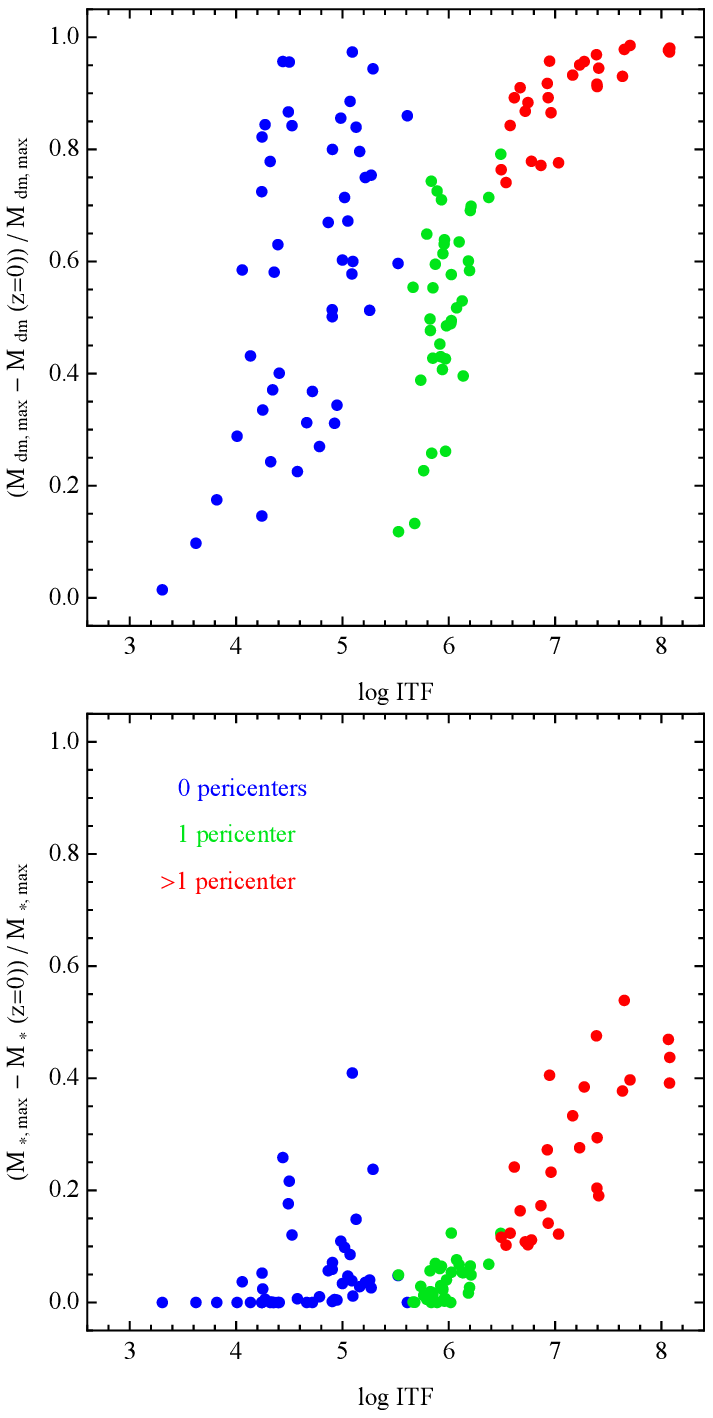}
\caption{Mass loss in galaxies associated with the cluster. Upper panel: the relative dark matter mass loss in
terms of the difference between the maximum and the present dark mass with respect to the maximum mass, as a function
of ITF. Lower panel: the same quantity for the stellar mass of the galaxies. The color coding is the same as
in Fig.~\ref{masses}.}
\label{tidalforcemass}
\end{figure}

In both panels of Fig.~\ref{masses}, the solid line indicates the equality between the dark and stellar masses. There
are seven data points for which the dark mass is smaller than the stellar mass. Six of these seven points belong to the
sample of strongly evolved galaxies and three of them correspond to the galaxies with Sublink catalog identification
numbers ID = 87, 92, and 101 whose mass and orbital evolution is shown in Fig.~\ref{evolution2}. In these cases, the
galaxies evolved on very tight orbits with many pericenters and lost a large fraction of their dark masses. Since the
dark matter distribution is usually more extended than the stars, it is more easily stripped by tidal forces. While
stellar mass is also lost, the effect is much milder than in the case of dark matter. We emphasize that these galaxies
are genuine cosmologically formed objects that were strongly dark matter dominated prior to their evolution in the
cluster. Their present low dark matter content is the result of strong tidal stripping by the tidal forces from the
cluster.

The mass history of the galaxies shown in Fig.~\ref{evolution_mass} and the upper rows of Figs.~\ref{evolution1} and
\ref{evolution2} is typical for all galaxies associated with the cluster. The dark mass (blue lines) reaches a maximum
value at some point in the past, after which it declines. In the case of evolved samples, this decline occurs when the
galaxy approaches the first pericenter of its orbit around the cluster. The maximum gas mass (green lines) usually
coincides in time with the time of the maximum dark mass. As the galaxy approaches the cluster center, the gas is
gradually depleted, but it continues to form stars so that the time when the maximum stellar mass is reached (red
lines) occurs much later than the maximum of the dark and gas content. At the same time, the gas is stripped by ram
pressure from the hot gas halo of the cluster, so the gas is completely lost soon after the first or second pericenter
passage. This process is reflected in the final (present) gas content of the galaxies: while only 16 out of 48 galaxies
(33\%) of the unevolved sample contain no gas at redshift $z=0$, this fraction rises to 25 out of 37 galaxies (67\%)
for the weakly evolved sample, and to 27 out of 27 objects (100\%) for the strongly evolved one.

The mass loss in the galaxies is further illustrated in Fig.~\ref{tidalforcemass}, where we show the relative mass loss
in dark matter (upper panel) and stars (lower panel) as a function of ITF. The values were calculated as the difference
between the maximum and the present mass divided by the maximum mass. We can see that, again, the weakly and strongly
evolved samples follow a branch of tight dependence on the ITF and are clearly separated within this branch. While the
galaxies of the weakly evolved sample lose between 10 and 80\% of their dark mass, the objects of the strongly evolved
one always lose more than 70\%. On the other hand, the mass loss in stars is much weaker, the galaxies with one
pericenter passage lose between 0 and 10\% of the stellar mass, while those with multiple pericenters lose between 10
and 55\% of the stars.

The results of this subsection concerning the mass loss are in good agreement with the studies of \citet{Smith2016}, who
used zoom-in cosmological simulations of clusters, and \citet{Niemiec2019}, who studied subhalo populations in clusters
of the original Illustris simulations. Both these studies found that, in general, dark matter is tidally stripped more
effectively than stars and that the galaxies continue to form stars after they are accreted by the cluster, which
makes the cluster galaxies less dominated by dark matter.

\begin{figure}
\centering
\includegraphics[width=7.1cm]{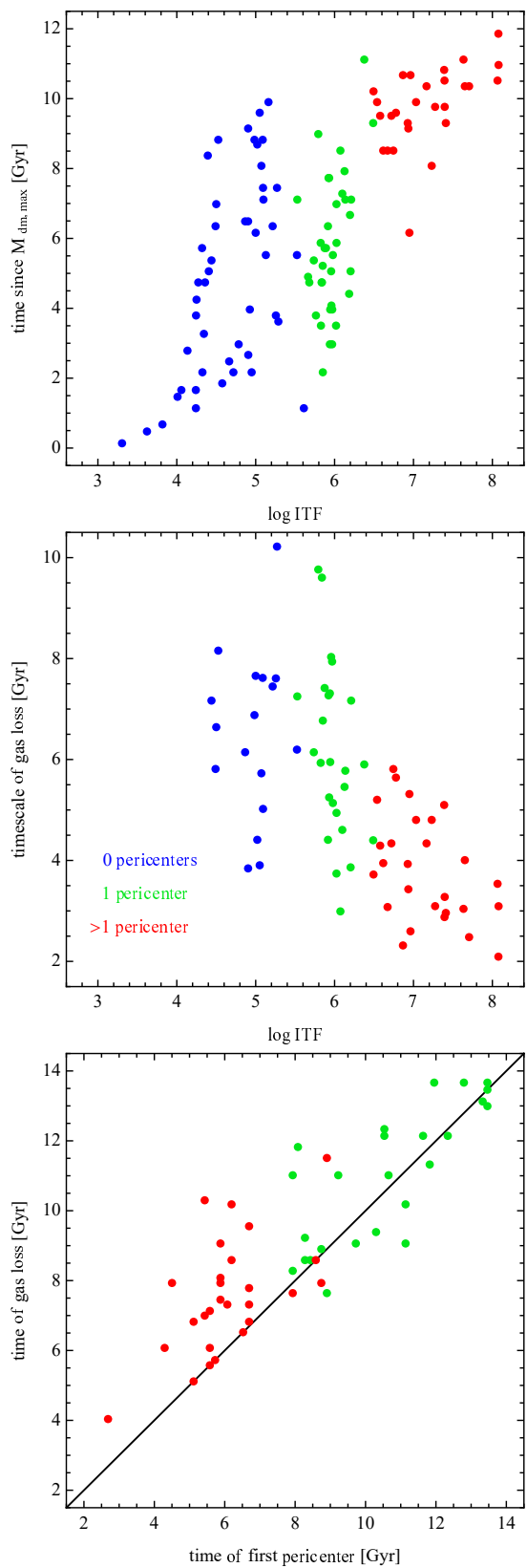}
\caption{Timescales of galaxy evolution in the cluster. Upper panel: time since the maximum dark
matter mass for all galaxies as a function of ITF. Middle panel: timescale of gas loss for the subsamples
of galaxies that do not contain any gas at $z=0$. Lower panel: time of gas loss as a function of the time
of the first pericenter passage for the galaxies that had at least one pericenter and lost all the gas. The diagonal
black line in the lower panel indicates the equality between these times. The color coding is the same as in
Fig.~\ref{masses}.}
\label{timescales}
\end{figure}

\subsection{Timescales of galaxy evolution}

Figure~\ref{timescales} illustrates the timescales of galaxy evolution in the cluster. In the upper panel, we plot the
times elapsed between the time when the galaxy reached its maximum dark matter mass and the present time as a function
of log ITF, which is the look-back time to the moment of maximum dark matter mass. The time ranges for the evolved
samples are again clearly separated and depend on the ITF reflecting the amount of time the galaxies spent in the
cluster. While the times for the weakly evolved sample (green points) cover the range between 2 and 11 Gyr, those for
the strongly evolved sample (red points) are in the range between 6 and 12 Gyr. This is expected, since galaxies with
more pericenters tend to enter the cluster and start losing mass earlier.

The middle panel of Fig.~\ref{timescales} shows the timescales of gas loss as a function of log ITF. Here, only
galaxies that lost all their gas until $z=0$ are shown. The timescales are measured as the time elapsed between the
time of the maximum gas mass the galaxy possessed during its lifetime and the time it lost all the gas. The dependence
on the strength of the tidal interaction is again very pronounced. The weakly evolved galaxies (green points) needed
between 3 and 10 Gyr to lose all their gas, while the strongly evolved objects (red points) need only between 2 and 6
Gyr to do so.

In the lower panel of Fig.~\ref{timescales}, we compare the time (in terms of the age of Universe) of the first
pericenter passage to the time when the galaxy loses all its gas. The comparison is obviously shown only for galaxies
that experienced at least one pericenter passage. In the case of strongly evolved galaxies (red points) the causal
connection between these two events is stronger: for all of these objects, except two, the gas is lost at or after the
first pericenter passage. For the weakly evolved sample (green points), this connection is not as tight, since for a
significant fraction of these galaxies the gas is lost even before the first pericenter. We note, however, that in most
cases of the evolved sample, the gas is not lost immediately at the first pericenter passage, and the process may take up
to 4 Gyr from the first pericenter (it sometimes coincides with the second pericenter passage).

\subsection{Evolution in color and metallicity}

\begin{figure}
\centering
\includegraphics[width=7.1cm]{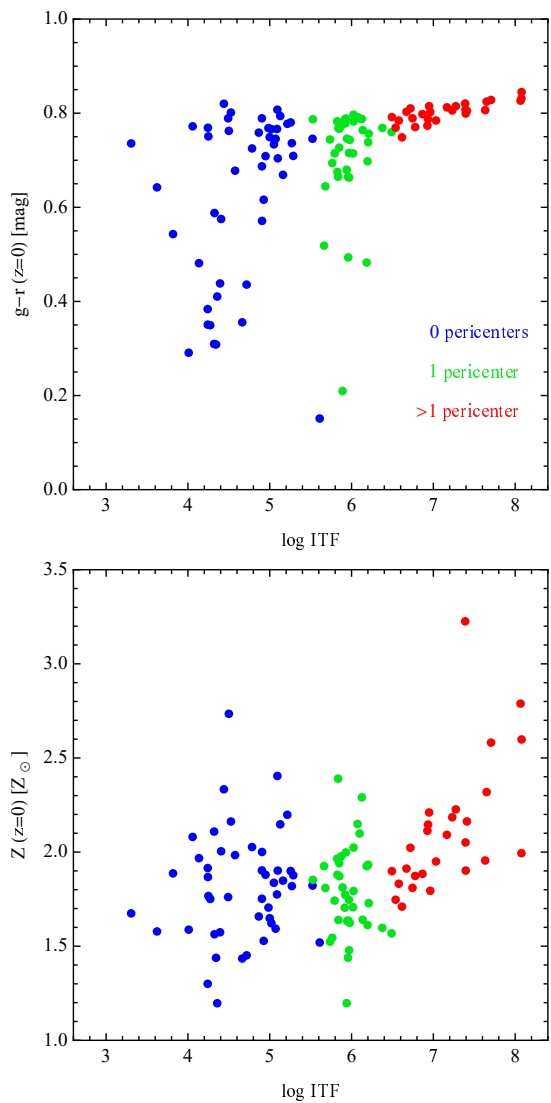}
\caption{Color and metallicity of galaxies in the cluster at $z=0$. Upper panel: color measure
$g-r$ as a function of ITF. Lower panel: metallicity $Z$ as a function of ITF. The color coding of the
points is the same as in Fig.~\ref{masses}.}
\label{tidalforcemc}
\end{figure}

\begin{figure}
\centering
\includegraphics[width=7.1cm]{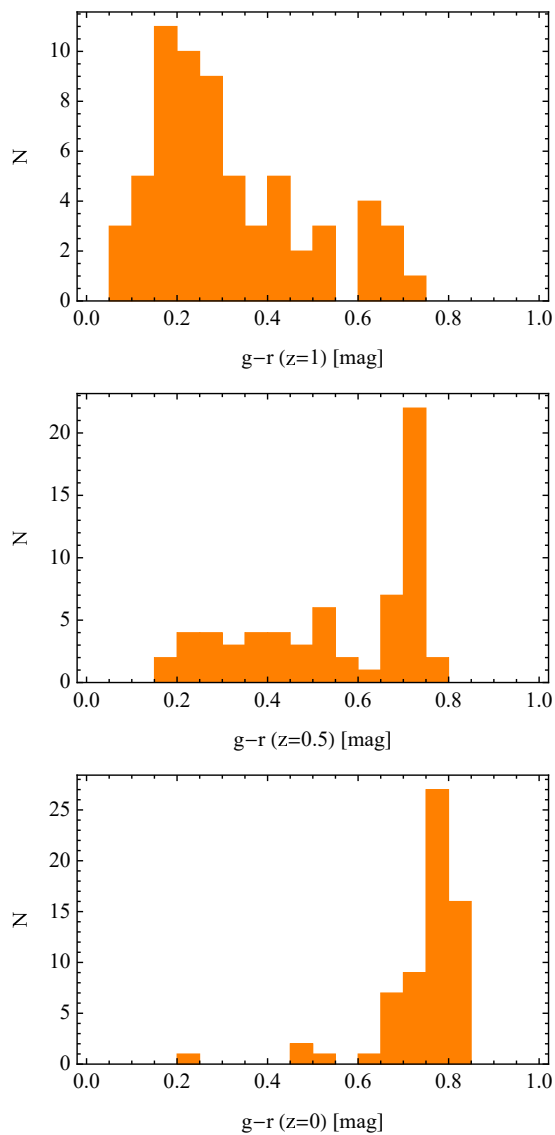}
\caption{Distributions of $g-r$ color for cluster galaxies at different redshifts. The three panels from top to
bottom show the distributions at redshift $z=1$, $z=0.5$ and $z=0$, respectively. Data for 64 galaxies from the weakly
and strongly evolved samples were included.}
\label{color}
\end{figure}

The gas loss discussed in the previous subsection has a great impact on the evolution of the cluster galaxies in terms
of star formation, which is suppressed. This environmental quenching leads to the change in galaxy color. IllustrisTNG
has been shown to provide a significant improvement over the original Illustris in reproducing galaxy colors in
comparison with the data from the Sloan Digital Sky Survey (SDSS) \citep{Nelson2018}. Thus, we may also expect to see some
signal in the color evolution for the cluster galaxies.

The Sublink catalogs of IllustrisTNG provide the magnitudes of galaxies in eight bands based on the summed-up
luminosities of all the stellar particles of the galaxy, from which we chose SDSS broad-band filters $g$ and $r$. We used
$g-r$ as a measure of galaxy color, in which case $g-r=0.6$ may be used as a value separating the blue ($g-r<0.6$) from
the red population ($g-r>0.6$). The present ($z=0$) values of the color for all our galaxies are plotted in the upper
panel of Fig.~\ref{tidalforcemc} as a function of ITF. We can see that while the infalling galaxies (blue dots) show a
wide range of colors, the galaxies become redder as we go to the evolved samples. Among the weakly evolved
galaxies (green points), there are still four with $g-r<0.6,$ but all the galaxies from the strongly evolved sample (red
points) are decidedly red with $g-r>0.7$.

In the lower panel of Fig.~\ref{tidalforcemc}, we also show the metallicities of the same sample of galaxies estimated
as the mass-weighted average metallicity (of all elements above He) of the stellar particles within twice the stellar
half-mass radius. Here, the dependence on the evolutionary stage of the galaxies is much weaker, but still the galaxies
from the strongly evolved sample tend to be more metal-rich. For each galaxy in the sample, we traced the evolution of
the color and metallicity in time. In the lower panels of Figs.~\ref{evolution1} and \ref{evolution2}, we plot examples
of such evolution for our six selected, strongly evolved galaxies. In all cases, the curves look similar: while the
metallicities increase rather fast early on and reach a constant level, the colors seem more related to the orbital
history. The galaxies remain blue until about the time of their first pericenter passage around the
cluster and soon after, when they lose all the gas, their color index rather quickly increases to above $g-r=0.6,$ and they
become red. The values of $g-r$ continue to grow slowly and reach $g-r=0.8$ by the present time. The $z=0$ values of
these quantities for the six galaxies are listed in the sixth and seventh column of Table~\ref{table}, respectively.

It is interesting to check when exactly the transition from the blue to red population takes place, that is when the
galaxies become red and dead as a result of their evolution in the cluster. To address this issue, in Fig.~\ref{color}
we show the distribution of the $g-r$ color at different redshifts for our sample of 64 evolved galaxies, meaning for
the weakly and strongly evolved samples combined. At redshift $z=1$ (upper panel), most of the galaxies are still blue,
with $g-r<0.6$, and at redshift $z=0.5$ (middle panel), the sample is divided equally into blue and red galaxies as 32
objects (exactly 50\%) have $g-r<0.6$ and the other half has the color larger than this value. At redshift $z=0$ (lower
panel), the population is strongly dominated by red galaxies. Therefore, redshift $z=0.5$ may be considered as the time
when the population of galaxies in the cluster becomes predominantly red. This is understandable in the context of the
orbital history of the galaxies in the evolved sample. As shown in the lower panel of Fig.~\ref{timescales}, at
redshift $z=1$ (corresponding to the age of the Universe $t=5.9$ Gyr) only a small fraction of galaxies that lost all
their gas by now have already experienced their first pericenter passage. Instead, at $z=0.5$ ($t=8.6$ Gyr) about half
of them did.

Although no similar comparison can be made for any real, observed cluster since different clusters are seen at
different redshifts, the evolution shown in Fig.~\ref{color} clearly explains the origin of the Butcher-Oemler effect,
namely that galaxy clusters at higher redshift contain a larger fraction of blue galaxies than the low-redshift ones
\citep{Butcher1978, Butcher1984, Rakos1995, Margoniner2001, Goto2003a}. If we treat the simulated cluster as
representative of the population of clusters at different redshifts, then given the large scatter in the fraction of
blue galaxies among clusters, different ways to measure color \citep{Goto2003a}, and the dependence on limiting
distance and cluster richness, the value of the blue fraction of the order of 0.5 at redshift $z=0.5$ found here seems
to be in reasonable agreement with observations \citep{Margoniner2001}.

\subsection{Morphological and kinematical evolution}

In addition to the tidal stripping that manifests itself in mass loss, the galaxies in the cluster also experience what
is sometimes referred to as tidal stirring. This process, proposed as a way to transform originally disky dwarf
galaxies into dwarf spheroidals in the context of the Local Group evolution, has been discussed in detail in a number
of studies \citep{Kazantzidis2011, Lokas2011b, Buck2019}. The repeated action of the tidal forces from a bigger host,
mostly during the pericenter passages, is supposed to reduce the ordered motions of the stars in the form of rotation
and increase the amount of random motion as well as thicken their disks, thus turning oblate rotating disks into
triaxial, prolate, or spherical objects dominated by random motions.

In order to quantify these changes in the case of our galaxies evolving in the cluster, we used the measures of the shape
and rotation provided by the Illustris team and calculated as described in \citet{Genel2015}. The shape is described
using the axis ratios of the stellar component determined within two stellar half-mass radii. They were estimated from
the eigenvalues of the mass tensor of the stellar mass inside this radius obtained by aligning each galaxy with its
principal axes and calculating three components ($i$=1,2,3): $M_i = (\Sigma_j m_j r^2_{j,i}/\Sigma_j m_j)^{1/2}$, where
$j$ enumerates over stellar particles, $r_{j,i}$ is the distance of stellar particle $j$ in the $i$ axis from the
center of the galaxy, and $m_j$ is its mass. The eigenvalues are sorted so that $M_1 < M_2 < M_3,$ therefore the
shortest to longest axis ratio is $c/a = M_1/M_3$, while the intermediate to longest is $b/a = M_2/M_3$. In order to
describe the shape of the galaxies with just one number, we also used the triaxiality parameter $T =
[1-(b/a)^2]/[1-(c/a)^2]$. With this parameter, one can classify the galaxy shapes into oblate ($T<1/3$), triaxial ($1/3
< T < 2/3$) and prolate ($T>2/3$).

\begin{figure}
\centering
\includegraphics[width=7.1cm]{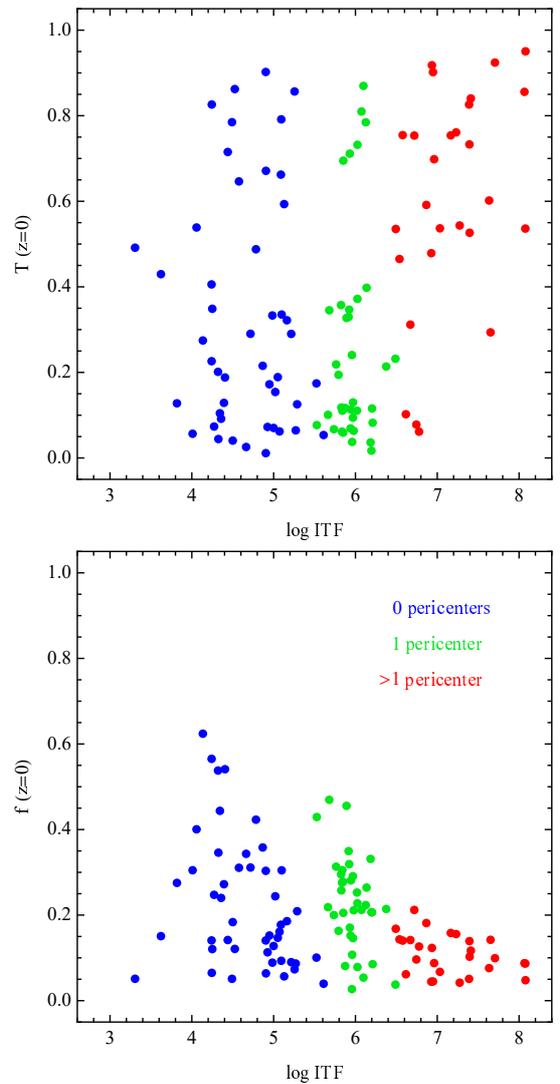}
\caption{Measures of shape and rotation for galaxies in the cluster at $z=0$. Upper panel: triaxiality
parameter $T$ as a function of ITF. Lower panel: rotation parameter $f$ as a function of ITF. The color coding
is the same as in Fig.~\ref{masses}.}
\label{tidalforcetf}
\end{figure}

\begin{figure}
\centering
\includegraphics[width=7.1cm]{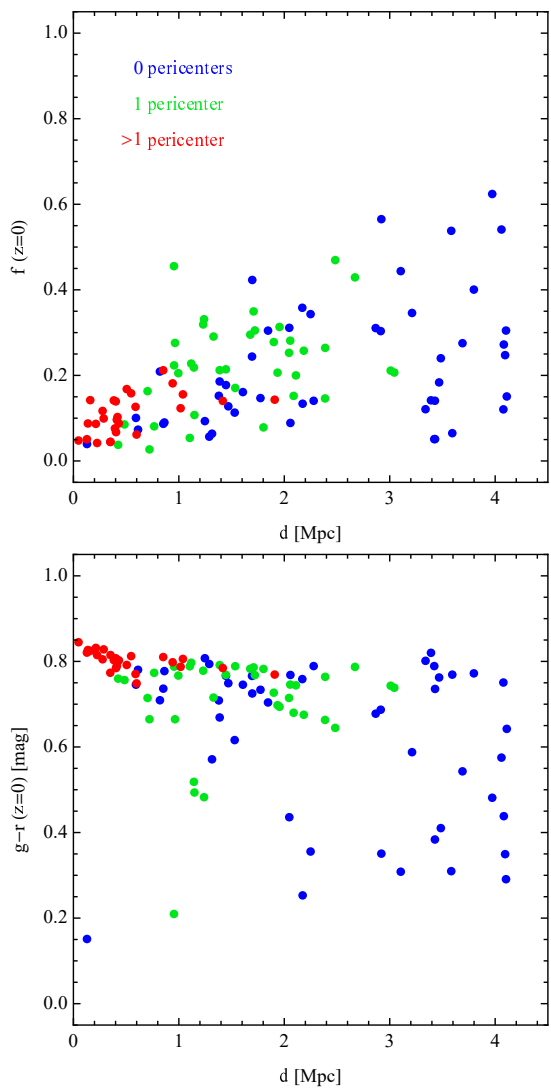}
\caption{Morphology-density relation for galaxies in the cluster. The upper and lower panel show,
respectively, the rotation parameter $f$ and the galaxy color $g-r$ at the present time as a function of the distance of
the galaxy from the cluster center. The color coding of the points is the same as in Fig.~\ref{masses}.}
\label{morden}
\end{figure}

The evolution of the axis ratios $c/a$, $b/a$ and the triaxiality parameter $T$ in time for our six selected galaxies
from the strongly evolved sample is shown in the third row of panels in Figs.~\ref{evolution1} and \ref{evolution2}. The
measurements of these parameters are restricted to subhalos with stellar mass $M_* > 3.4 \times 10^8$ M$_\odot$
within twice the stellar half-mass radius, and at least 100 stars so they do not extend so far into the past as the
measurements of the mass and orbital history. The final ($z=0$) values of these parameters for these six galaxies are
listed in the eighth, ninth, and tenth column of Table~\ref{table}. The $z=0$ values of triaxiality for all our
galaxies are plotted in the upper panel of Fig.~\ref{tidalforcetf} as a function of log ITF. We can see that while the
weakly evolved sample (green points) is dominated by oblate objects (majority of galaxies have $T<0.5$), prolate galaxies are more numerous (majority of galaxies have $T>0.5$) in the
strongly evolved sample (red points).

In order to quantify the amount of rotation support in the galaxies, we used the rotation parameter $f$ defined as the
fractional mass of all stars with circularity parameter $\epsilon > 0.7,$ where $\epsilon=J_z/J(E)$. Here $J_z$ is the
specific angular momentum of the star along the angular momentum of the galaxy, while $J(E)$ is the maximum angular
momentum of the stellar particles at positions between 50 before and 50 after the particle in question in a list where
the stellar particles are sorted by their binding energy. The evolution of the rotation parameter $f$ for our selection
of six galaxies is shown in the fourth row of the panels in Figs.~\ref{evolution1} and \ref{evolution2}. The final
values of this parameter (at $z=0$) are given in the 11th column of Table~\ref{table}.

The final snapshot values of $f$ for all galaxies are plotted in the lower panel of Fig.~\ref{tidalforcetf} as a
function of log ITF. As in the case of shape parameters, there is a systematic trend of $f$ evolution when we go from
the least to the most evolved samples, to such an extent that there is less rotation support in the more evolved
galaxies. While the galaxies from our weakly evolved sample (green points) take values from a wide range of $f$ between
0 and 0.5, the galaxies from the strongly evolved sample (red points) almost all have $f < 0.2$. We note that $f > 0.2$
is a reasonable threshold to define late-type or disky galaxies in Illustris \citep{Peschken2019, Peschken2020}, so $f
< 0.2$ means that our evolved galaxies are no longer disks.

The results on the shape and kinematics evolution of this subsection combined with the color changes discussed
in the previous subsection indicate that indeed as the galaxies orbit the cluster, preferably on tight orbits with many
pericenter passages, they change their shapes from disky to triaxial or prolate, and their rotation is suppressed and
replaced by random motions while they become increasingly red. In Fig.~\ref{morden}, we again plot the rotation
parameters $f$ and colors $g-r$ of the galaxies at the present time, but as a function of their distance from the
cluster center. Since the distance is the indicator of cluster density, the Figure illustrates the presence of the
morphology-density relation in the simulated cluster. The galaxies with little rotation support and red colors, which
would be classified as early type by observers, are predominantly found near the cluster center, while those with a
significant amount of rotation and blue colors (late type) are mostly present in the outskirts of the cluster. The
trend is thus similar to the morphology-density (or morphology-radius) relation found in real clusters
\citep{Dressler1980, Dressler1984, Goto2003b, Postman2005}.

\begin{figure*}
\centering
\includegraphics[width=5cm]{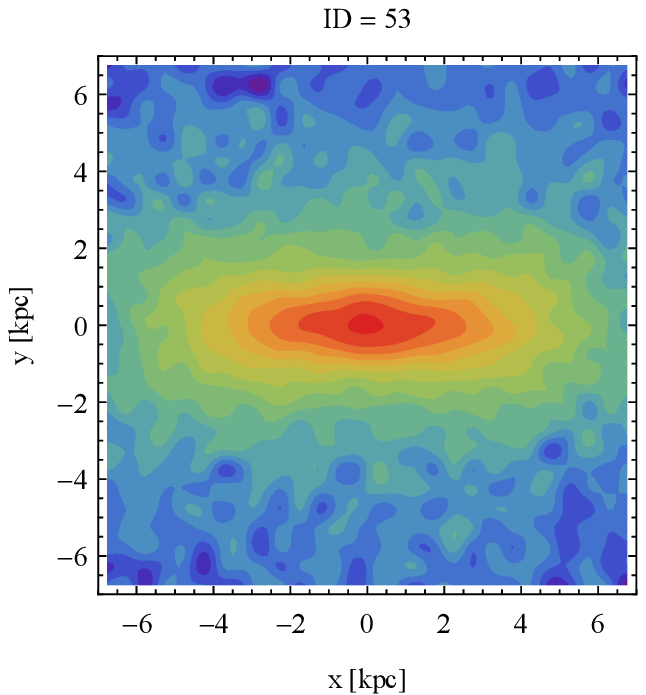}
\includegraphics[width=5cm]{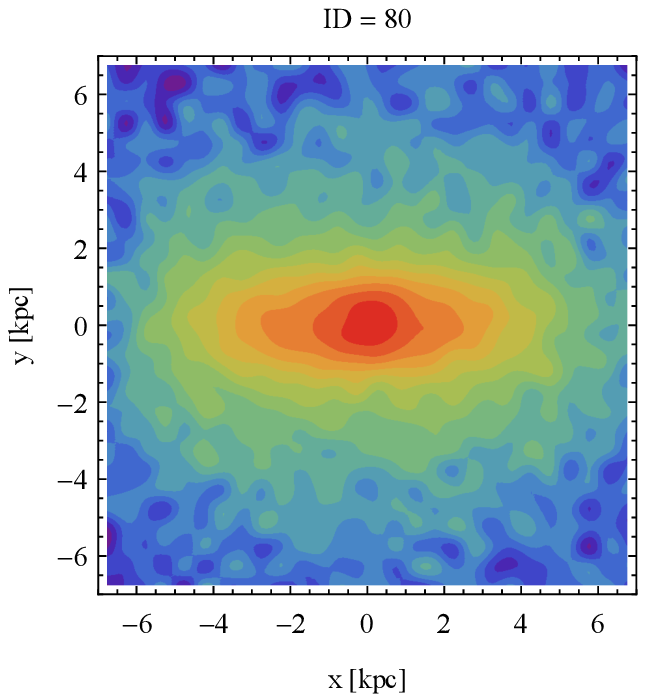}
\includegraphics[width=5cm]{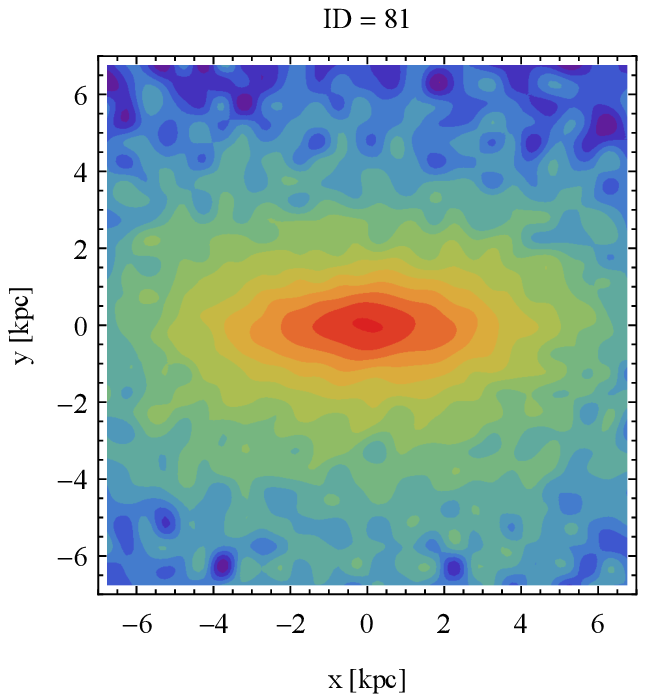} \\
\vspace{0.2cm}
\includegraphics[width=5cm]{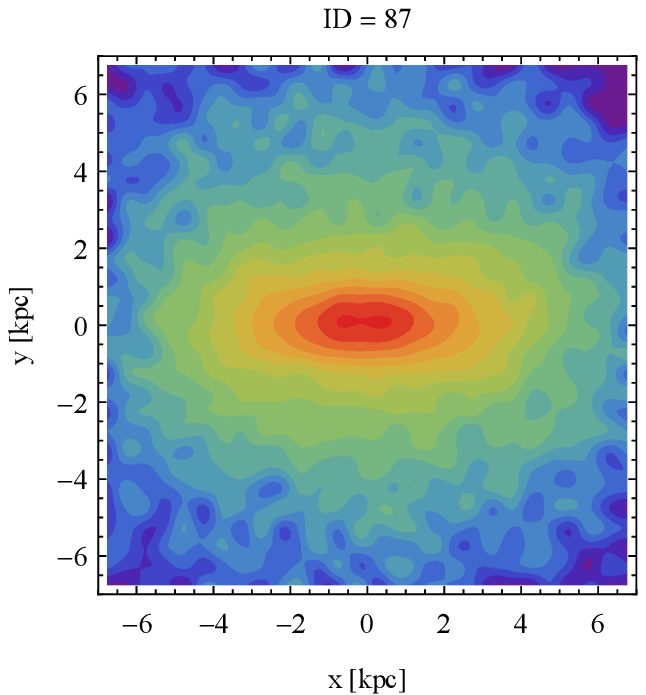}
\includegraphics[width=5cm]{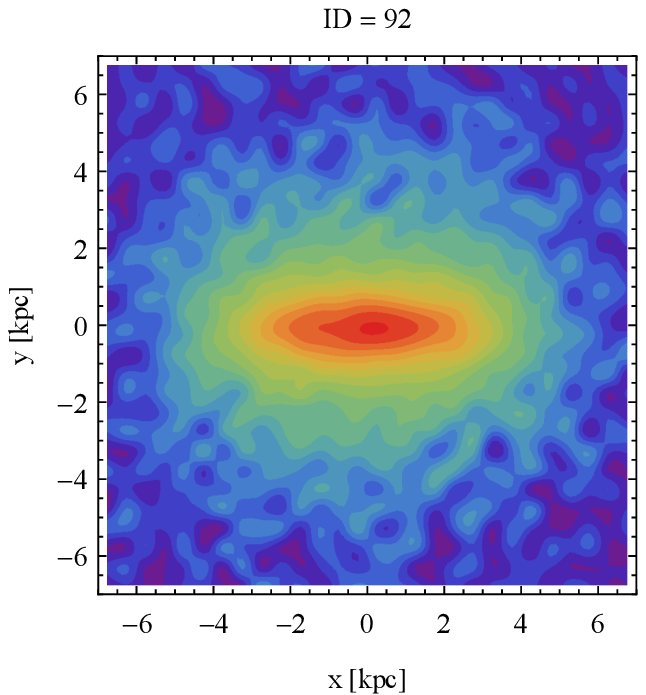}
\includegraphics[width=5cm]{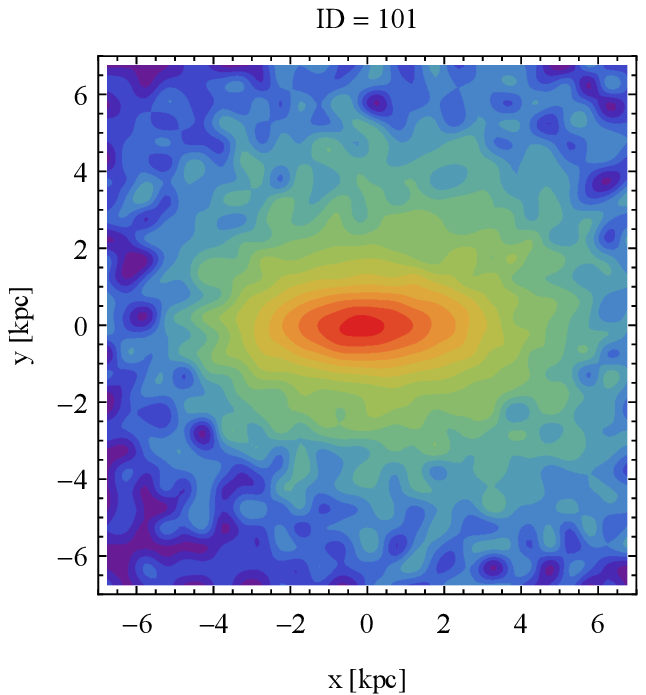} \\
\vspace{0.2cm}
\hspace{0.62cm}
\includegraphics[width=4.3cm]{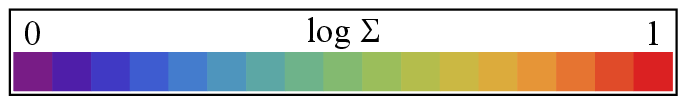}\\
\vspace{0.2cm}
\caption{Surface density distributions of the stellar component viewed face-on for galaxies identified as tidally
induced bars. The surface density was normalized to the central maximum value in each case, and the contours are equally
spaced in $\log \Sigma$.}
\label{surden}
\end{figure*}

\subsection{Tidally induced bars}

The evolution of the shape and kinematics parameters shown in the third and fourth rows of the panels in
Figs.~\ref{evolution1} and \ref{evolution2} reveals that all these objects were in the past oblate rotating disks,
meaning they had their rotation parameters $f>0.2$ and their triaxiality parameters $T<1/3$. Later on, at some point,
the rotation parameters started to decrease and the triaxiality parameters increased. For these six galaxies, the time
of this transition clearly coincides with one of the first pericenter passages of the galaxy on its orbit around the
cluster. With time, $f$ decreased down to almost zero while $T$ reached values above 2/3. This means that the disks turned
into prolate, slowly rotating spheroids, with properties similar to tidally induced bars.

To verify this, we extracted, from the last simulation snapshot, the cutouts containing the stellar particle data for
these galaxies and rotated the stellar positions to align the $x$, $y,$ and $z$ coordinates with the principal axes
using the same procedure as was used to determine the axis ratio. The coordinate system was chosen so that the $x$ axis
is along the longest axis of the stellar component and the $y$ and $z$ axes are along the intermediate and shortest
ones, respectively. The projections of the stellar mass density along the shortest $z$ axis for the six galaxies from
Figs.~\ref{evolution1} and \ref{evolution2} are shown in Fig.~\ref{surden}. They correspond to the face-on views of the
stellar component and clearly reveal the presence of elongated shapes along the $x$ axis, which is characteristic of bars.

In order to further confirm that these are indeed tidally induced bars, we calculated the measure of the bar
strength as the $m=2$ mode of the Fourier decomposition of the surface density distribution of stellar particles
projected along the short axis: $A_m (R) = | \Sigma_j m_j \exp(i m \theta_j) |/\Sigma_j m_j$. Here, $\theta_j$ is the
azimuthal angle of the $j$th star, $m_j$ is its mass and the sum is up to the number of particles in a given radial
bin. The radius $R$ is the standard radius in cylindrical coordinates in the $xy$ plane, $R = (x^2 + y^2)^{1/2}$. The
results of such measurements within two stellar half-mass radii are shown as a function of time in the fourth
row of panels in Figs.~\ref{evolution1} and \ref{evolution2} together with the rotation parameter $f$. We can see that
before accretion the galaxies were normal disks with a lot of rotation and no bar, as the $A_2$ parameter remains close
to zero. After one of the pericenter passages $A_2$ starts to increase (while the rotation decreases) signifying the
formation of a tidally induced bar for all galaxies.

\begin{figure}
\centering
\includegraphics[width=7.1cm]{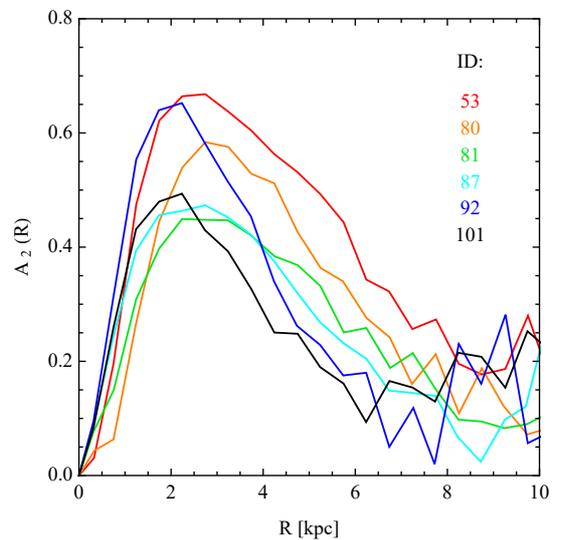}
\caption{Profiles of bar mode $A_2 (R)$ for galaxies identified as tidally induced bars. The measurements were done
in bins of $\Delta R = 0.5$ kpc containing at least 10 stars.}
\label{a2profiles}
\end{figure}

The profiles of the bar mode as a function of radius, $A_2(R)$, are plotted in
Fig.~\ref{a2profiles} for the six
galaxies. The profiles show a behavior characteristic of galactic bars: they increase
to a maximum value and then decrease to almost zero. The maximum values of the profiles, $A_{2,{\rm max}}$, can be
used as a measure of the bar strength and are listed in the 12th column of Table~\ref{table}. All these values are
above 0.4, signifying rather strong bars, and they are well within the range of values for bars in observed galaxies
\citep{Diaz2016}. The profiles in Fig.~\ref{a2profiles} can be also used to estimate the length of the bars as the
radius $R,$ where $A_2 (R)$ drops to half the maximum value. These lengths turn out to be of the order of 4-6 kpc, in
agreement with the visual impression from Fig.~\ref{surden} and with the observed values
\citep{Aguerri2005, Diaz2016, Font2017}.

\begin{figure}
\centering
\includegraphics[width=4.4cm]{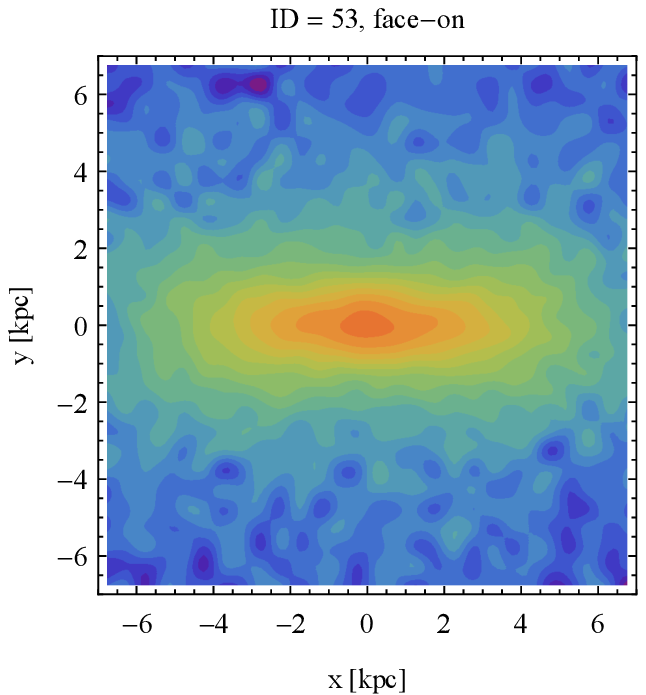}
\includegraphics[width=4.4cm]{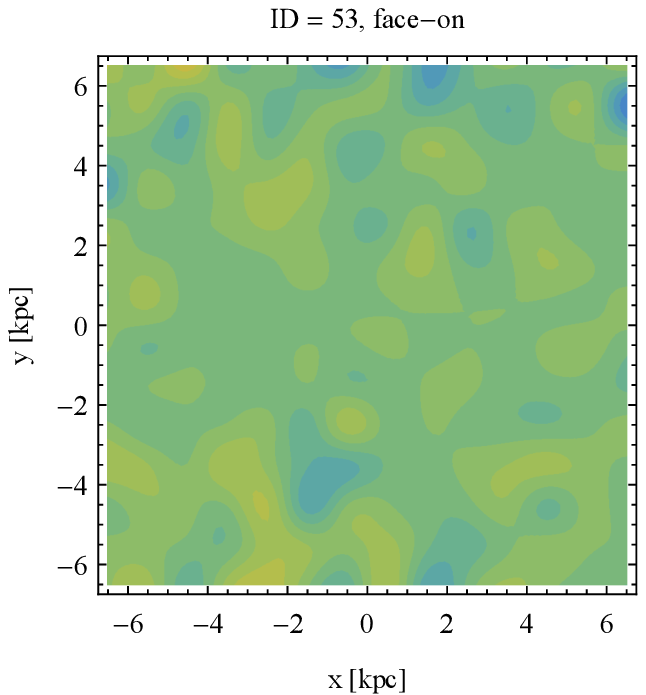}\\
\vspace{0.3cm}
\includegraphics[width=4.4cm]{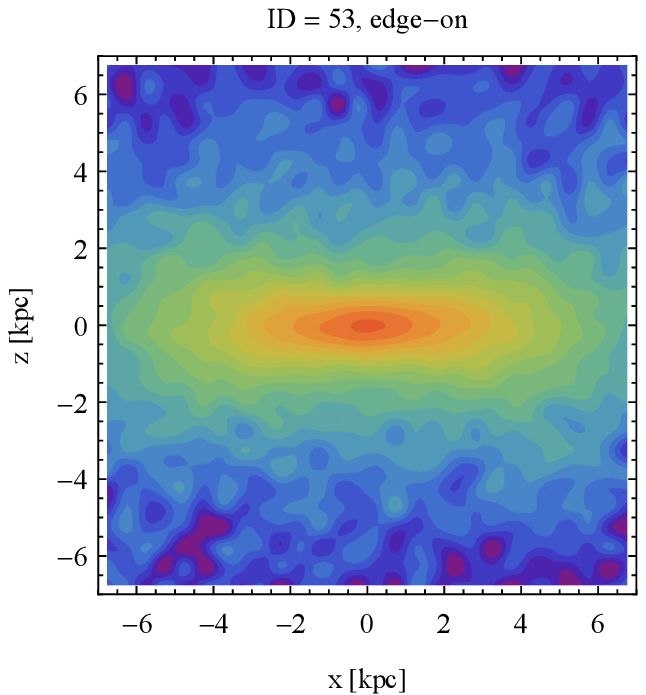}
\includegraphics[width=4.4cm]{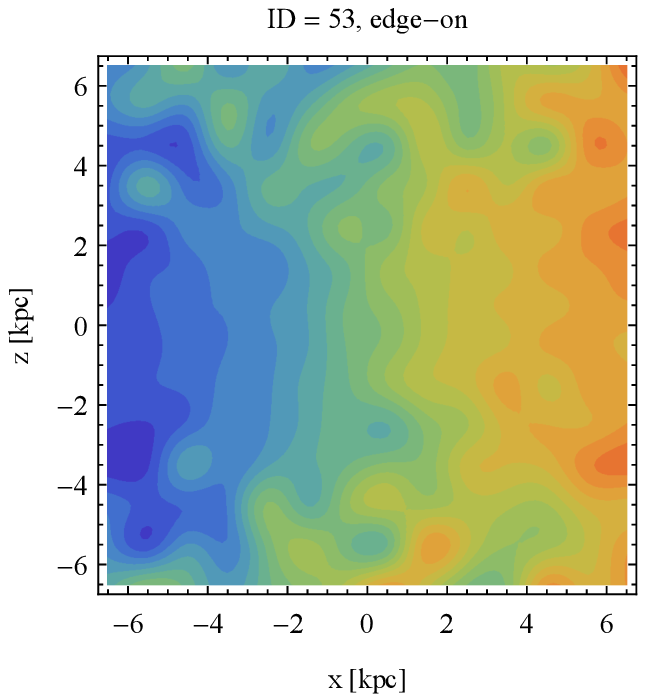}\\
\vspace{0.3cm}
\includegraphics[width=4.4cm]{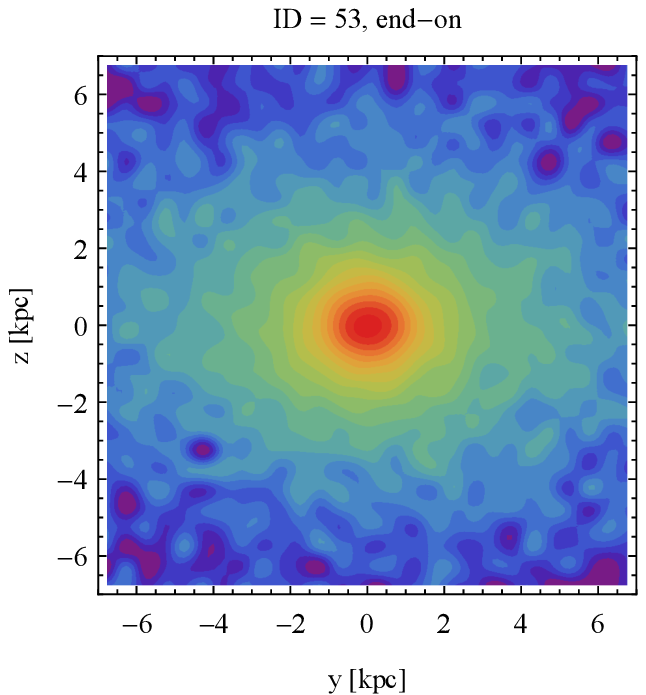}
\includegraphics[width=4.4cm]{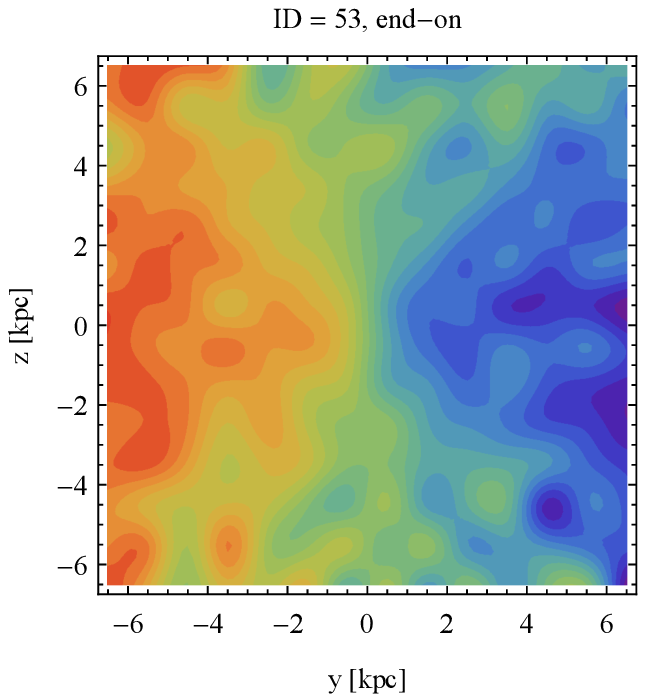}\\
\vspace{0.2cm}
\hspace{0.54cm}
\includegraphics[width=3.79cm]{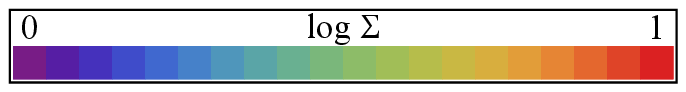}
\hspace{0.53cm}
\includegraphics[width=3.79cm]{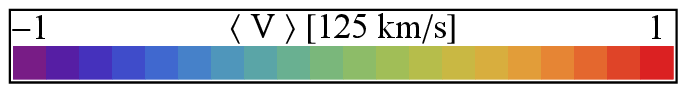}\\
\vspace{0.3cm}
\caption{Surface density distribution (left column) and the line-of-sight velocity field (right column)
of the stars for the galaxy with ID=53. Rows show, respectively, the face-on, edge-on, and end-on view. The surface density
and velocity were normalized to the maximum values measured in the end-on view (along the bar).}
\label{denvel}
\end{figure}

Since the six galaxies we identified as tidally induced bars are prolate spheroids rather than bars embedded in disks,
we also looked at the velocity distribution of their stars. One example, for our strongest tidally induced
bar (ID=53), is shown in Fig.~\ref{denvel}. In all cases, the velocity field measured in projection along the three
principal axes of the stellar component shows features characteristic of bars \citep{Athanassoula2002}. Namely, there
is a strong rotation signal when the galaxy is seen edge-on and end-on, while there is no systematic rotation in the
face-on view, so the galaxies rotate around their shortest axis. This means that these objects are indeed bars and not
prolate spheroids with prolate rotation (around the long axis) of the kind studied by \citet{Ebrova2017}.

We also estimated the pattern speeds $\Omega_{\rm b}$ of these tidally induced bars. Since the simulation
outputs are spaced quite widely in time, they cannot be measured directly by the change of the position of the bar
major axis between outputs and we had to use the kinematic method proposed by \citet{Tremaine1984}. Although the
velocity field of the bars seems quite regular, judging by the example in Fig.~\ref{denvel}, these
measurements are rather noisy due to the low resolution of the simulated galaxies. Still, for all the bars, the method
yields values of the order of 10-20 km s$^{-1}$ kpc$^{-1}$, with high uncertainty, but similar to those obtained by
\citet{Peschken2019} for the population of tidally induced and all bars in Illustris. These values are, however, on the
low end of the range typically measured in real barred galaxies \citep{Font2017}.

The six galaxies whose properties are illustrated in Figs.~\ref{evolution1}-\ref{evolution2} and
\ref{surden}-\ref{a2profiles} were selected as the most convincing examples of bars induced by the galaxy interaction
with the cluster. These galaxies did not experience any significant merger with another subhalo since their accretion
onto the cluster and the abrupt change in their morphology and kinematics coincides with one of their first pericenter
passages. However, it is difficult to prove beyond reasonable doubt that their transformation is caused by their
interaction with the cluster, as some other small accretion events or interactions (not recorded in the merger trees)
are still possible, and the nature of bar instability may be to some extent stochastic \citep{Sellwood2009}.

The picture painted by the analysis of these six examples is, however, strongly supported by the more statistical
approach we presented in the previous subsection. The differences in the distributions of $T$ and $f$ values for the
weakly and strongly evolved samples in Fig.~\ref{tidalforcetf} demonstrate that indeed on average as the galaxies orbit
the cluster, their rotation tends to decrease, and their prolateness to increase, which supports the interpretation in
terms of tidally induced bars.

Among the sample of strongly evolved galaxies in Fig.~\ref{tidalforcetf}, in addition to the six objects
already described in detail, there are seven other galaxies with $T>2/3,$ which suggests a bar-like shape. One of them
(ID=11) is prolate but also quite spherical with axis ratios $b/a$ and $c/a$ of the order of 0.8, and so it cannot be
considered a bar. The remaining six (ID=34, 45, 58, 66, 96, 105) are genuine bars, but their morphology
and kinematics changes occur before or after the pericenter passage, or their stellar distribution was already a
bit elongated before the interaction so they could be enhanced by the cluster but not necessarily induced by it. We
note that there is no correlation between the strength of the bar, as measured by $A_{2, {\rm max}}$, and the ITF,
because although strong tidal interaction can induce a bar, a repeated action of tidal forces can also reduce its
strength later on by making the galaxy more spherical.

The remaining 14 galaxies of the strongly evolved sample have intermediate or low values of $T$ and one may wonder why
they did not form bars despite their significant interaction with the cluster. These cases have very different
histories, but their failure to produce strong bars by $z=0$ can be traced to one or a combination of the following
reasons: (1) they had bars earlier, but they were weakened by $z=0$; (2) they did not possess enough rotation initially;
(3) their pericenter distances were rather large, so the strength of the interaction was smaller; (4) their orbits in
the cluster were retrograde, which is known to suppress bar formation \citep{Lokas2015, Lokas2018, Peschken2019}. In
particular, the three galaxies of this sample that remain oblate with $T \sim 0.1$ at $z=0$ with ID = 21, 90, 124 fall
into the categories (3), (2)-(3), and (2)-(3)-(4), respectively.

Finally, we note that out of six galaxies with $T > 2/3$ in the weakly evolved sample (green points in
Fig.~\ref{tidalforcetf}), four are quite spherical with high axis ratios, and two are more bar-like, but in all cases the
transition to prolate shapes is not related to the pericenter passage, which is usually quite recent, but rather to
earlier mergers and interactions. On the other hand, the unevolved sample (blue points in Fig.~\ref{tidalforcetf}),
which also contains a significant number of prolate objects, includes bona fide bars corresponding to the points with
the highest $T>0.8$ (ID = 44, 55, 91). The formation of these bars is not related to the fact that the galaxies are
infalling into the cluster and took place earlier.

\begin{figure*}[ht!]
\centering
\includegraphics[width=18.2cm]{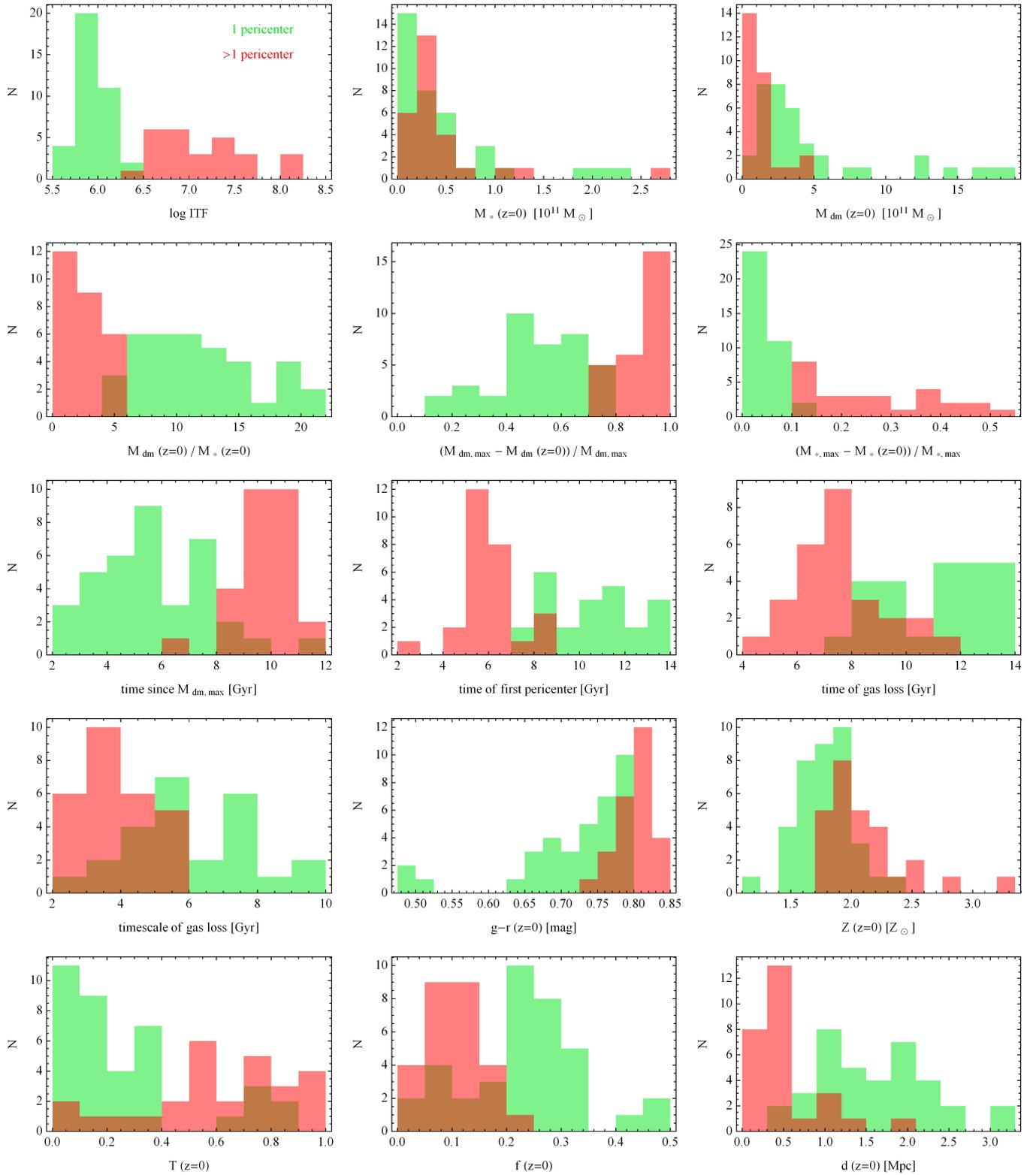}
\caption{Histograms of distributions of different properties in the weakly evolved (one pericenter, green) and
strongly evolved (more than one pericenter, red) sample of galaxies. The panels summarize the results shown in
Figs.~\ref{masses}-\ref{tidalforcemc} and \ref{tidalforcetf}-\ref{morden} by binning the data instead of plotting them
as points, with $N$ corresponding in each case to the number of galaxies in a given bin.}
\label{histograms}
\end{figure*}

\section{Summary and discussion}

We studied the evolution of galaxies in the most massive cluster of the Illustris TNG-100 simulation. The galaxies were
characterized by deriving their orbital history, mass evolution, and changes in their color and metallicity as well as in
their morphology and kinematics. We divided our sample of galaxies into three subsamples according to their orbital
evolution: the unevolved group of infalling objects, the weakly evolved galaxies that passed just one pericenter, and
the strongly evolved ones that experienced more than one pericenter up to now. Our results are summarized in
Fig.~\ref{histograms}, where we plot the distributions of galaxies with different properties for our weakly and strongly
evolved samples of galaxies. The properties considered are the same as those discussed in detail in Section~2 and shown
in Figs.~\ref{masses}-\ref{tidalforcemc} and \ref{tidalforcetf}-\ref{morden}. We do not show the distributions for the
unevolved sample as these galaxies only served as a reference. The data in Fig.~\ref{histograms} were binned to
emphasize the difference between the subsamples.

Figure~\ref{histograms} clearly demonstrates that the distributions are distinctly different for the two samples. This
means that as the galaxies evolve in the cluster, their properties change significantly. In particular, we find that in
comparison to the weakly evolved sample, the galaxies of the strongly evolved sample experienced much larger integrated
tidal force, their first pericenter took place earlier and they reached their maximum dark mass a longer time ago. The
dark matter fraction of these galaxies is now much lower, and they lost a bigger fraction of their dark and stellar
mass. They lost their gas earlier and on a shorter timescale. Finally, they became redder and more metal rich and lost
more rotation while their shapes evolved into more prolate. In addition, a significant fraction of the
strongly affected galaxies formed tidally induced bars or had them enhanced as a result of their interaction with the
cluster.

The results presented here demonstrate that galaxies are indeed affected by their evolution in clusters. In the lower
right panel of Fig.~\ref{histograms}, we show the distributions of the present ($z=0$) distances of the
galaxies from the cluster center for both subsamples. As expected, the galaxies from the strongly evolved sample
concentrate near the cluster center and are all within the cluster virial radius (2 Mpc), while the radial distribution
of the galaxies from the weakly evolved sample is more extended. The three lower panels of
Fig.~\ref{histograms} together with the histograms for color and metallicity again confirm that the morphology-density
relation is reproduced in the simulated cluster: the galaxies in the denser region are more prolate or spherical, have
less rotation, and are redder than the galaxies in the outskirts of the cluster.

While describing the evolution of different properties of the galaxies in the cluster, we used the concept of the
integrated tidal force, the total tidal force from the cluster experienced by a given galaxy during its entire history.
The concept, originally used in \citet{Lokas2011a} to study the properties of dwarf galaxies evolving around the Milky
Way in controlled simulations, also proved to be very useful here, in the study of galaxy evolution in the cluster
environment, and it allowed us to isolate the effect of the cluster from the global cosmological evolution. We found that
different galaxy properties such as mass loss, evolution timescales, and morphological properties depend very strongly
on ITF. We thus confirm the essential similarity of the tidal evolution processes in environments such as small groups
and massive clusters of galaxies and demonstrate that in both cases they lead to the changes in galactic properties
that manifest themselves in the form of the morphology-density relation.

Many of the galaxies in the strongly evolved sample turned out to possess or indeed "be" tidally induced bars. They
"are" tidally induced bars in a sense that their stellar component is essentially a bar-like spheroid rather than
a bar embedded in a disk. We identified six such tidally induced bars among the 27 galaxies of the strongly evolved
sample. Their tidal origin was confirmed by the fact that the time of their morphological and kinematic transformation
coincides with one of their first pericenter passages around the cluster. Six more galaxies of this sample probably had
their bars enhanced rather than induced by their interaction with the cluster. We therefore confirm the picture
proposed in \citet{Lokas2016} that a significant fraction of barred galaxies in clusters can be of tidal origin. While
\citet{Lokas2016} used an idealized approach of controlled simulations of a single galaxy orbiting a cluster, we
corroborate that the scenario is also valid in the cosmological context of IllustrisTNG.

If this scenario is true, we should expect to find a higher fraction of barred galaxies toward the centers of
clusters. Unfortunately, there is still no convincing observational evidence in favor or against this hypothesis. Some
studies \citep{Thompson1981, Andersen1996, Barazza2009} found a higher concentration of barred galaxies in the
cluster center, while others \citep{Mendez2010, Lansbury2014, Lokas2016} found no (or very weak) evidence for such a trend, pointing instead to other factors contributing to the bar fraction, such as stellar mass rather than
environment \citep{Cervantes2015}.

Recently, \citet{Peschken2019} studied tidally induced bars in galaxies from the original Illustris simulation. Their
sample included the whole population of galaxies, without the restriction to cluster members, with a similar resolution
threshold as adopted here. However, their selection of galaxies included only late-type objects, with rotation parameter
$f > 0.2$ at the present time, that is galaxies that, while forming a bar, retained a significant amount of rotation and
their disky character. Instead, almost all the galaxies of our strongly evolved sample and all the tidally induced or
enhanced bars had $f < 0.2$ as a result of their evolution in the cluster, which means that they would not be included
in the late-type sample similar to the one used by \citet{Peschken2019}. This may explain why their sample
included mainly fly-by type of interactions with other galaxies, and why the fraction of barred galaxies they found in
Illustris was lower than observed among real galaxies.

It would be interesting to compare the properties of the tidally induced bars formed in the cosmological
context to those produced in controlled simulations such as those of \citet{Gerin1990},
\citet{Miwa1998}, \citet{Lang2014}, or \citet{Lokas2018}. One of the predictions of such simulations is that
tidally induced bars should have lower pattern speeds than the bars formed in isolation by disk instability.
Such a comparison is rather difficult due to different resolutions and the lack of correspondence between cases. Still,
it was attempted by \citet{Peschken2019} who, however, found no statistically significant difference between
the pattern speeds of tidally induced bars and the whole population of bars. This is at least partly due to the low
accuracy of such measurements, which we also experienced here.

In addition to our analysis of the mass loss and morphological evolution of the galaxies, we also looked into the
evolution of their gas component. We found that only 1/3 of infalling galaxies are gas free, and the fraction increases
to 2/3 and 1 for the weakly and strongly evolved samples, respectively. This means that the cluster environment is
extremely effective in stripping the galaxies of their gas. We also determined that the strongly evolved galaxies lost
their gas much faster, over 2-6 Gyr, while the weakly evolved ones needed 3-10 Gyr to do so. In most cases, the gas is
not lost immediately after the first pericenter, but it may take up to 4 Gyr from that point, in contrast to the results of
\citet{Lotz2019}, who found much shorter timescales. Our results are, however, in very good agreement with those of
\citet{Jung2018}, who studied the mechanisms of gas depletion of galaxies in simulated clusters. They also found 1/3 of
galaxies to be gas poor before entering the cluster as a result of pre-processing in groups, a significant fraction
that lost their gas prior to the first pericenter passage, and some that retained their gas after the first pericenter.

In agreement with \citet{Rosas2020}, who studied the population of massive barred galaxies in IllustrisTNG, we found
that the fraction of bars is larger among gas-poor galaxies, since bars are mostly found among our strongly evolved,
gas-free sample. However, in our case this is not related to the bar quenching but rather to the processes of gas
stripping and bar formation happening concurrently as the galaxies evolve in the cluster. The two processes may be related, however, because gas loss is supposed to enhance bar formation, since strong bars form more easily in galaxies
with lower gas fractions \citep{Athanassoula2013, Lokas2020}.

\begin{acknowledgements}
I am grateful to the anonymous referee for useful comments and to the IllustrisTNG team for making their
simulations publicly available. This work was supported in part by the Polish National Science Center under grant
2013/10/A/ST9/00023.
\end{acknowledgements}

\end{document}